\documentclass[aps,prl,twocolumn,longbibliography,superscriptaddress,amsmath]
{revtex4-1}

\usepackage[colorlinks, urlcolor=blue,linkcolor=blue, anchorcolor=blue, citecolor=blue]{hyperref}
\usepackage[english]{babel}
\usepackage[utf8]{inputenc}
\usepackage{multirow}
\usepackage{array}
\usepackage{amsthm}
\usepackage{mathtools}
\usepackage{physics}
\usepackage{xcolor}
\usepackage{graphicx}
\usepackage{bm}
\usepackage{soul}
\usepackage{adjustbox}
\usepackage{placeins}
\usepackage[T1]{fontenc}
\usepackage{lipsum}
\usepackage{csquotes}
\usepackage{float}
\usepackage{booktabs}

\newcommand{\Teff}{$T_{\rm eff}$}
\newcommand{\Tavg}{$\langle T\rangle$}
\newcommand{\Tin}{$\langle T_{\rm in}\rangle$}
\newcommand{\Tf}{$\langle T_{\rm f}\rangle$}
\newcommand{\Nf}{n_{\!f}}
\newcommand{\muB}{\mu_{\rm B}}

\def\bea{\begin{eqnarray}}      \def\eea{\end{eqnarray}}
\def\pd{\partial} 
\def\dd{{\rm d}}
\def\sNN{s_{_{\rm NN}}}

\def\Im{\hbox{Im}}

\def\gsim{\mathrel{\rlap{\lower0.25em\hbox{$\sim$}}\raise0.2em\hbox{$>$}}} 
\def\lsim{\mathrel{\rlap{\lower0.25em\hbox{$\sim$}}\raise0.2em\hbox{$<$}}}
\def\lg{\mathrel{\rlap{\lower0.25em\hbox{$>$}}\raise0.25em\hbox{$<$}}}
\def\gl{\mathrel{\rlap{\lower0.25em\hbox{$<$}}\raise0.25em\hbox{$>$}}}

\begin{document}

\title{Virtual Photons Shed Light on the Early Temperature of Dense QCD Matter}

\author{Jessica Churchill}
 \affiliation{%
   Department of Physics, McGill University, 
   3600 University Street, Montreal, QC H3A 2T8, Canada}
\author{Lipei Du}
\affiliation{%
    Department of Physics, McGill University, 
    3600 University Street, Montreal, QC H3A 2T8, Canada}
\author{Charles Gale}
 \affiliation{%
   Department of Physics, McGill University, 
   3600 University Street, Montreal, QC H3A 2T8, Canada}
\author{Greg Jackson}
 \affiliation{%
   Institute for Nuclear Theory, Box 351550, 
   University of Washington, Seattle, WA 98195-1550, United States}
\affiliation{%
   SUBATECH, 
   Nantes Universit\'e, IMT Atlantique, IN2P3/CNRS,
4 rue Alfred Kastler, La Chantrerie BP 20722, 44307 Nantes, France}
\author{Sangyong Jeon}
 \affiliation{%
   Department of Physics, McGill University, 
   3600 University Street, Montreal, QC H3A 2T8, Canada}
\date{\today}

\begin{abstract}
    Dileptons produced during heavy-ion collisions represent a unique probe of the QCD phase diagram, and convey information about the state of the strongly interacting system at the moment their preceding off-shell photon is created. In this study, we compute thermal dilepton yields from Au+Au collisions performed at different beam energies, employing a (3+1)-dimensional dynamic framework combined with emission rates accurate at next-to-leading order in perturbation theory and which include baryon chemical potential dependencies. By comparing the effective temperature extracted from the thermal dilepton invariant mass spectrum with the average temperature of the fluid, we offer a robust quantitative validation of dileptons as effective probe of the early quark-gluon plasma stage.
\end{abstract}

\maketitle

{\it \textbf{Introduction.}---}
To clarify the many-body properties of quantum chromodynamics (QCD),
like its emergent phases and their boundaries, 
remains a chief objective in nuclear physics~\cite{Braun-Munzinger:2008szb}. 
Relativistic nuclear collisions as performed and studied in terrestrial laboratories constitute the only means to explore 
the properties of QCD matter under extreme but controlled conditions~\cite{Bzdak:2019pkr}. 
The data generated by these events can then be used to 
inform our understanding  of QCD, 
and to push further the
extent of our knowledge.
These explorations come with significant challenges. 
For one, 
systems generated in heavy-ion collisions are 
highly dynamic and short-lived, evolving over mere yoctoseconds~\cite{Shuryak:2014zxa}. 
The trajectory 
of the strongly interacting system across 
the QCD phase diagram, from 
birth of the 
quark-gluon plasma
(QGP)
to emergence of confined hadrons, is complex and comprises 
various stages. 
To decipher the properties of the transient nuclear matter, 
an advanced multistage framework is required, 
and the resulting predictions must be compared to sophisticated many-body experimental observables~\cite{Bernhard:2019bmu,JETSCAPE:2020shq,Nijs:2020ors,Heffernan:2023utr,Niemi:2015qia,Hirvonen:2022xfv,Parkkila:2021yha,Parkkila:2021tqq}. 

The fact that hadrons interact strongly with the medium 
makes them mostly sensitive to the late stages 
of the evolution; 
this complicates the extraction of information about the early stage of QGP evolution from hadronic measurements. 
Electromagnetic probes are not handicapped in the same way:
Real and virtual photons get generated continuously throughout the entire collision evolution, but unlike hadrons, they remain unaltered by the strong interactions once 
emitted~\cite{Shuryak:1978ij,Kajantie:1981wg,McLerran1984,Hwa:1985xg,Kajantie:1986dh}. 
This distinctive feature 
has made 
them exceptional tools for investigating the early-stage QGP~\cite{Peitzmann:2001mz,Salabura:2020tou,Geurts:2022xmk}. 
Among electromagnetic probes, lepton pairs (dileptons) resulting from the decay of virtual photons, are especially useful as their invariant mass renders their spectrum impervious to flow effects, unlike that of real photons  which can be 
altered by Doppler shifts~\cite{vanHees:2011vb,Shen:2013vja} \footnote{See however the recent study of Ref. \cite{Paquet:2023bdx}.}.
Thus, dileptons are usually considered a reliable thermometer for assessing the properties of the hot and dense QCD medium \cite{Rapp:2014hha,HADES:2019auv}, even if their emission rate is suppressed over that of real photons by a factor of the fine structure constant, $\alpha_{\rm em}$.  
Nonetheless, it's important to acknowledge that dileptons are generated at various stages of the collision fireball \cite{Salabura:2020tou,Geurts:2022xmk}, 
wherein the temperature undergoes significant fluctuations both in 
space and time. 
Therefore, 
there remains a need 
to establish a clear connection between 
the effective temperatures derived from dileptons and the underlying physical properties of the medium. 
That connection, which  requires
delicate modeling, is the purpose of this paper. 

Specifically, our goal is to investigate the fidelity of dilepton spectra as ``thermometers'' of the excited partonic medium formed in nucleus-nucleus collisions at energies of the Relativistic Heavy-Ion Collider (RHIC), from the Beam Energy Scan (BES) regime to the top RHIC energy: $\sqrt{s_{_{\rm NN}}} = 7.7-200$~GeV. 
In order to compute the dilepton yields originating 
from the baryon-charged QCD medium 
existing at collision energies on the low side of this range, 
we utilise the dilepton emission rate at next-to-leading order (NLO) 
with nonzero baryon chemical potential.
We compare values of the temperature extracted from dilepton spectra in the intermediate invariant mass region (IMR),
 $1$~GeV~${\leq M \leq 3}$~GeV~\footnote{%
 As found in Refs.~\cite{Rapp2013,Vujanovic2013,Vujanovic2016}, the QGP source dominates over dileptons from a hadronic medium for $M > 1.1$~GeV.
},
with the ``true'' values occurring in the (3+1)-dimensional dissipative hydrodynamics that is tuned to reproduce the hadronic measurements. 
In so doing, we
establish a reliable connection between the effective temperatures extracted from dileptons and the fundamental physical properties of the QCD medium.

{\it \textbf{Thermal dilepton radiation.}---}
The yield of emitted thermal dileptons, 
can be obtained from the time and volume integrated
rate of a QGP that has attained {\em local} thermal equilibrium.
We let $T(X)\,$, $\muB(X)$ and $u^\mu(X)$
describe the local temperature, baryon chemical potential and 
flow velocity 
of the plasma
respectively,
where $X=(t,\bm x)$ is a spacetime coordinate.
Conservation of energy, momentum, and baryon current
dictate the hydrodynamic evolution of the system~\cite{Heinz:2013th,Denicol2018,Du:2019obx},
with viscous corrections controlled by transport coefficients~\cite{Denicol2018,Shen:2020jwv}, 
and as constrained by an equation of state~\cite{Monnai2019}. 

In finite-temperature field theory, 
the fully differential rate is related to the in-medium self-energy of the photon, 
$\Pi^{\mu \nu}$~\cite{Weldon:1990iw,Gale:1990pn},
which is calculated as a function of the dilepton's energy, $\omega\,$,
and momentum, $\bm k\,$, in the local rest frame.
As such, they are spacetime dependent:
\begin{equation*}
\omega(X) = K_\mu u^\mu(X) \, , \quad 
k(X)=\sqrt{\omega(X)^2 -M^2} \, ,
\end{equation*}
where
$K^\mu = (M_\perp \cosh y, \bm k_\perp, M_\perp \sinh y)$
is the (measured) four-momentum of the dilepton, in the lab frame,
with the $z$-direction aligned parallel to the 
axis of the colliding nuclei, $M_\perp \equiv \sqrt{M^2 + k_\perp^2}$ is the transverse mass and, $y$ being the rapidity. The yield with respect to $M$ and $y$ can be
 expressed as
\begin{align}
\frac{\dd N}{\dd M \, \dd y} 
&=
\frac{\alpha_{\rm em}^2}{3\pi^3  M}
\ \bigg\{  \sum_{i=1}^{\Nf} Q_i^2 \bigg\}B \Big( \frac{m_\ell^2}{M^2} \Big)
\,\nonumber\\
&\times  
\int \dd^2 \bm k_\perp \,
\int \dd^4 X \ \frac{
\Im \, \Pi_\mu^{\ \mu} \big( \omega(X), k(X) \big) }{
\exp \big({\omega(X)/T(X)}\big)-1
} \, ,
\label{eq:yield}
\end{align}
where 
the quark charge-fractions are $Q_i$ (in units of the electron charge), 
and the kinematic factor to produce the pair of leptons is 
$B(x) \equiv (1 + 2x) \sqrt{ 1 -4x }$ if $x<\frac14\,$, otherwise $B(x)\equiv0\,$. 
Three light flavours are assumed, i.e.\ $\Nf=3\,$, implying that
$\sum_i Q_i^2 \to \frac23\,$.
And since we focus on the IMR, 
the lepton masses can be set to zero ($m_\ell = 0$) and $B\simeq 1$
in Eq.~\eqref{eq:yield}.

\begin{figure}[t]
\begin{align}
\Pi^\text{LO}\ \, &:
\vcenter{\hbox{
\includegraphics[scale=.5]{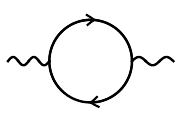}
}} & {\cal O}(1)
\nonumber\\
\Pi^\text{NLO} &:
\vcenter{\hbox{
\includegraphics[scale=.5]{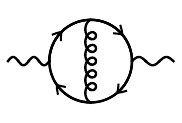}
}} 
\vcenter{\hbox{
\includegraphics[scale=.5]{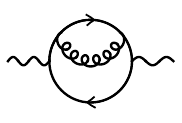}
}}
\vcenter{\hbox{
\includegraphics[scale=.5]{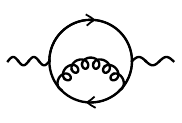}
}} \hspace{-1cm}& {\cal O}(\alpha_s)
\nonumber\\
\!\! \Delta
\Pi^\text{LPM} &:
\vcenter{\hbox{ 
\includegraphics[scale=.5]{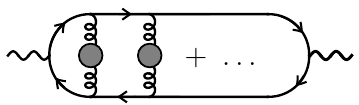}
}} 
& {\cal O}(\alpha_s^{i}) \, , \ i \geq 2
\nonumber
\end{align}
\vskip -2mm
\caption{\label{fig:pQCD} 
The perturbative diagrams included in our evaluation.
For the LPM class of ladder diagrams, $\vcenter{\hbox{\protect\includegraphics[scale=.55]{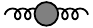}}}$
indicates that rungs are screened with HTL gluon self-energies
while it should be understood that the valence quarks
are evaluated at their asymptotic thermal mass.
}
\end{figure}

The rates can be derived 
from the imaginary part of the 
retarded photon self-energy~\cite{Gale:1990pn,Kapusta:2006pm,Laine:2016hma}. 
We evaluate the 
QCD corrections to the dilepton emission rate
in perturbation theory
and include non-zero values of the baryon chemical potential 
$\muB$~\cite{Churchill:2023vpt}. 
For intermediate masses $M$,
it becomes necessary to interpolate between two regimes~\cite{Ghisoiu2014},
namely: $i$) for $M \gsim T$, and $ii$) for $M \lsim \sqrt{\alpha_s}\,T$.
In the former case, strict order-by-order perturbation theory 
can be used, see $\Pi^{\rm LO}$ and $\Pi^{\rm NLO}$ in Fig.~\ref{fig:pQCD}, but this approach breaks down as 
the mass $M$ becomes `parametrically' 
small~\footnote{%
{
The overlap between the regimes cannot be neglected because $\alpha_s$ is not asymptotically small in practice.}
}.
The upshot is that 
arbitrary orders in $\alpha_s$ are needed to describe both
screening via hard thermal loop (HTL) insertions~\cite{Braaten:1989mz},
in addition to the  
Landau-Pomeranchuck-Migdal (LPM) effect~\cite{Aurenche2002,Aurenche:2002pc,Arnold2001ba,Arnold2001ms,Aurenche2002}.
This resummation can be performed rigorously for $M\to 0$,
involving ladder diagrams as shown in Fig.~\ref{fig:pQCD},
but only includes an approximate form of 
the strict one and two-loop self energy
[when formally re-expanded to ${\cal O}(1)$ and ${\cal O}(\alpha_s)$
respectively].
Therefore, we marry regime $i$) with regime $ii$) by using the full LO and NLO expressions and only keeping the higher order parts of the LPM result~\footnote{%
To be clear, we are only using the `leading-order' LPM spectral function. Subsequent QCD corrections in this limit have been computed~\cite{Ghiglieri2013,Ghiglieri2014}. 
}.
The higher order LPM corrections are necessary to obtain a finite result when $M=0\,$, and near this point they
serve to compensate the remnant of an unphysical log divergence
in $\Im\, \Pi^{\rm NLO}$ for $M\to 0^\pm$~\cite{Baier1988,Gabellini1989,Altherr1989,Kapusta1991,Baier1991}.
Far away from the lightcone, the LPM corrections are formally not justified, but remain negligible next to the LO and NLO parts~\cite{Laine2013vpa,Laine:2013vma,Jackson2019a}.

With these considerations in mind, we 
adopt the full resummed
spectral function~\cite{Jackson2019}, defined by
\begin{align}
\Im\,\Pi^{\rm NLO}_{\rm resummed}
&=
  \Im\big[\, \Pi^{\rm LO} 
+ \Pi^{\rm NLO}
+ \Delta \Pi^{\rm LPM} \,\big] \, .
\label{eq:resummed}
\end{align}
The formal power counting in $\alpha_s$ is indicated in Fig.~\ref{fig:pQCD},
although it should be noted that collinear singularities reorganise the naive interpretation of certain diagrams \cite{Arnold2001ms}. 
Each ingredient in \eqref{eq:resummed} needs
to be evaluated numerically, for which the details
can be found in Ref.~\cite{Churchill:2023vpt}.
For the QCD coupling, we use the fixed value $\alpha_s=0.3$
which is motivated in Sec.~\ref{sec:running}
of the Supplementary Material~\cite{suppl} (where we also comment on the overall accuracy of perturbation theory in the IMR).

\begin{figure}[!tb]
\begin{center}
\includegraphics[width=0.9 \linewidth]{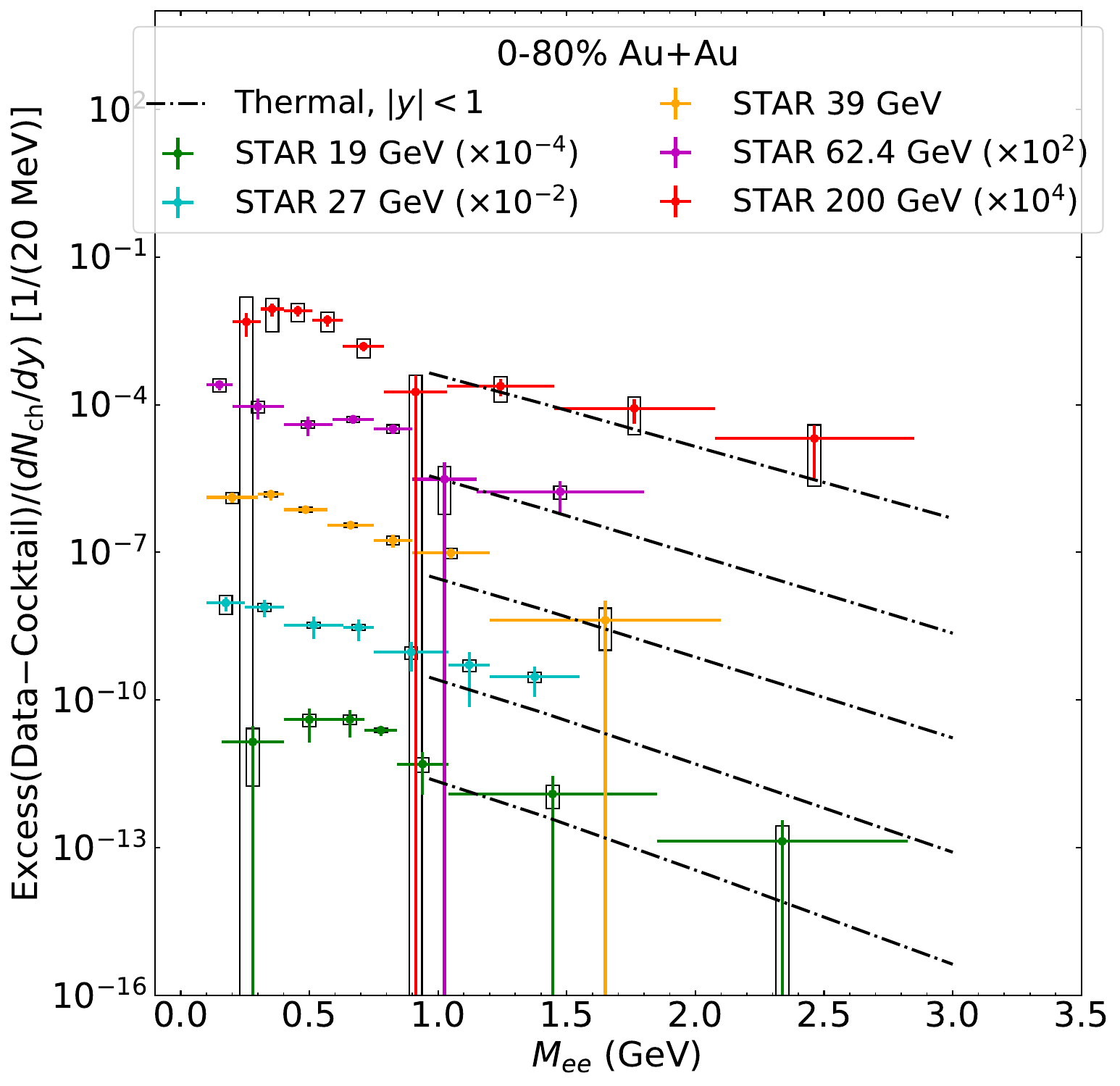}
\vspace{-2mm}
\caption{%
    Dielectron excess mass spectra within rapidity $|y|{\,<\,}1$, normalized by midrapidity charged hadron multiplicity $\dd N_{\rm ch}/\dd y$, for 0--80\% Au+Au collisions at $\sqrt{s_{_{\rm NN}}}=19,\,27,\,39,\,62.4,\,200\,$GeV. The markers with error bars and open boxes indicate the STAR experimental measurements with statistical and systematic uncertainties, while the dot-dashed lines represent model calculations.
    }
    \label{fig:dilepton_yields}
\end{center}
\end{figure}

The integration in Eq.~\eqref{eq:yield} is performed over a 
(3+1)-dimensional fluid dynamical evolution~\cite{dileptoncode,iebe}, 
 with four-volume element $\dd^4 X = \tau \dd \tau \, \dd \eta_s \, \dd^2 \bm x_\perp$ where $\tau$ is the longitudinal proper time
and $\eta_s$ is the spacetime rapidity, 
specifically calibrated to reproduce the {\em hadronic} data measured at the energies discussed in this paper~\cite{Du:2023gnv,Churchill:2023vpt}. 
The hydrodynamical framework includes both shear viscosity and baryon diffusion, but bulk viscosity is neglected and
the dilepton emission rates themselves are not corrected for viscous effects~\footnote{%
Doing so for the LPM contribution is challenging~\cite{Hauksson:2017udm}, and left for potential future work.
}. 
We integrate $k_\perp$ 
over a range sufficient  for comparison with the acceptance-corrected excess spectra measured by the STAR Collaboration~\cite{STAR:2013pwb,STAR2015,STAR:2015zal,STAR:2023wta}.
Since this study focuses on thermal dileptons originating from the QGP, 
we specifically consider fluid cells with temperatures
exceeding the freeze-out line as established in Ref.~\cite{Cleymans:2005xv}. 
This demarcation
closely aligns with the chemical freeze-out line determined by the STAR Collaboration~\cite{STAR:2017sal}, 
and we attribute the thermal dileptons emitted from the fluid cells below this line to contributions from hadronic matter.
In Fig.~\ref{fig:dilepton_yields}, one can see that the calculated thermal signal is in fact in quantitative agreement with measurements performed by the STAR Collaboration~\cite{STAR:2013pwb,STAR2015,STAR:2015zal,STAR:2023wta}, once the background contributions~\footnote{Those are labeled ``cocktail'' by the experimental collaboration. They are lepton pairs coming from the Drell-Yan process, from semi-leptonic decays of open flavor mesons, and from radiative decays of final-state hadrons.} have been subtracted. 
While $\muB$ increases at lower beam energies, the corresponding dependence in the emission rate
leaves little imprint of this on the dilepton spectra 
themselves 
because a majority of the QGP fluid 
satisfies $\muB/T\lsim~3$ within $|\eta_s|<1$ even at a beam energy of $7.7$~GeV.

{\it \textbf{Probing the early QGP temperature.}---} 
In our longer companion paper~\cite{Churchill:2023vpt}, 
we establish the effectiveness of the temperature extraction 
method using the thermal dilepton spectra within the IMR 
as a proxy for the temperatures of the fluid cells. 
That region of dilepton invariant mass is chosen to 
highlight the signal coming from the QGP phase, as 
lower invariant masses are known to receive important contributions from reactions involving composite hadrons~\cite{Rapp:2009yu}. 
Furthermore, the perturbative scheme in Eq.~\eqref{eq:resummed}
is well suited for this mass range 
with an estimated theoretical uncertainty 
comfortably below $\sim 10\%$.

The approximate large-$M$ behaviour of the rate, ${\rm d}\Gamma/{\rm d}M \sim (MT)^{3/2}\exp(-M/T)\,$, motivates defining an effective temperature \Teff{} by assuming that the integrated spectrum ${\rm d}N/{\rm d}M$ follows a similar functional form in the IMR. 
We perform an analysis where we derive \Teff{} from dilepton spectra at each time step, 
and then examine how it relates to the evolving hydrodynamic 
temperatures as functions of proper time. 
Figure~\ref{fig:temp_extraction_tau} in the Supplementary Material \cite{suppl} clearly illustrates that \Teff{} closely tracks the mean hydrodynamic temperature \Tavg{}
as a function of proper time, 
reflecting the cooling of the expanding QGP fireball. The close alignment between \Teff{} and \Tavg{} 
provides another compelling validation of the temperature extraction method,
although we see that \Teff{} is consistently above \Tavg{}
throughout the evolution.

\begin{figure}[tbp]
\begin{center}
\includegraphics[width= 0.8\linewidth]{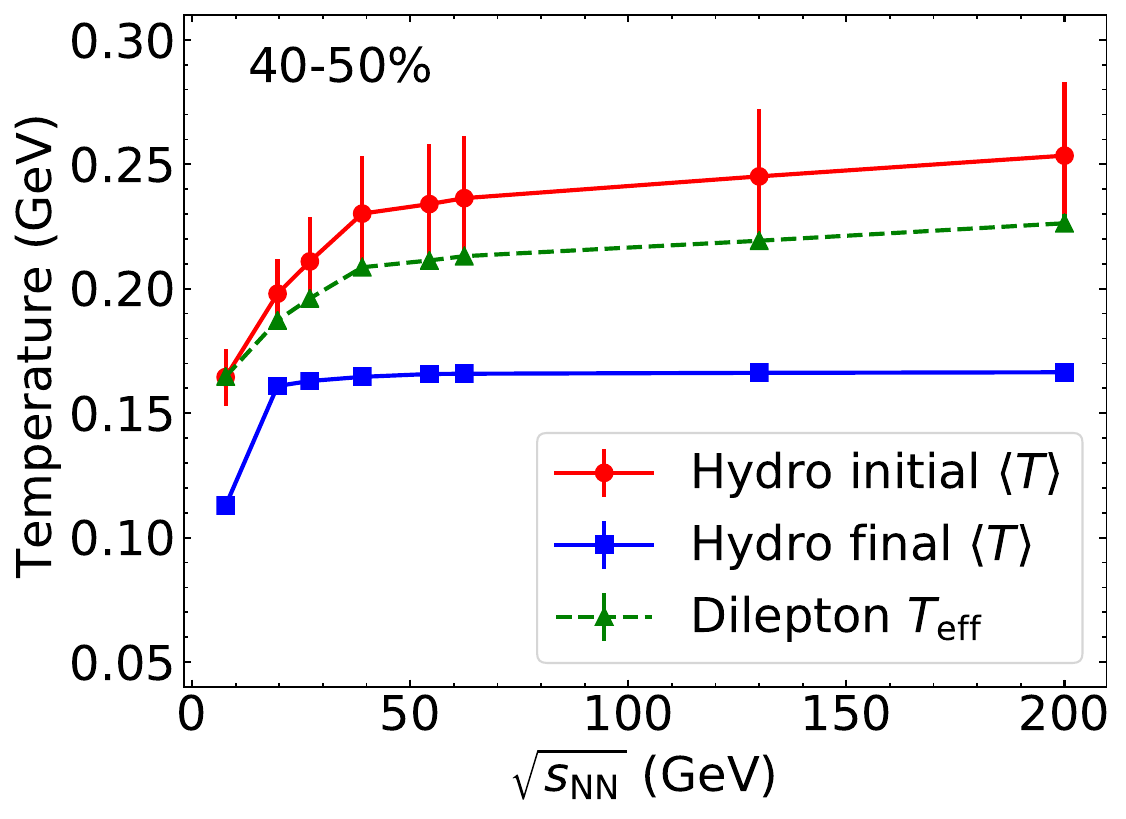}
\vspace{-2mm}
\caption{%
        Initial temperatures (red dots) and final temperatures (blue squares) of the hydrodynamic evolution, along with effective temperatures derived from dilepton spectra (green triangles) for 40--50\% Au+Au collisions, presented as a function of beam energy. The error bars associated with the hydrodynamic temperatures indicate standard deviations, while the error bars of the effective temperatures represent uncertainties resulting from the fitting procedure in the temperature extraction method.
        }
    \label{fig:initial_final_temp}
\end{center}
\end{figure}

Building upon these validations, we apply the same temperature extraction method to determine \Teff{} from the spacetime-integrated thermal dilepton spectra seen in Fig.~\ref{fig:dilepton_yields}.
To make the most of \Teff{} as a reliable thermometer for the QGP fireball, 
we further investigate its connection with the initial temperatures \Tin, 
marking the beginning of the hydrodynamic expansion, 
and final freeze-out temperatures \Tf{} 
across the eight beam energies.
This is illustrated in Fig.~\ref{fig:initial_final_temp} 
(for 40--50\% centrality),
which includes the initial temperature variations originating from both 
short-range lumpy fluctuations,
and long-range gradients on the system scale.
(The latter is the dominant factor, which 
shrinks with lower beam energies 
because of the lower temperatures at the fireball's center.)
Figure~\ref{fig:initial_final_temp} shows that the final freeze-out temperature 
remains relatively stable down to
$\sqrt{s_{_{\rm NN}}}=\,$19~GeV but experiences a sudden drop at $7.7$~GeV. 
This decline is due to the significantly higher $\muB/T$ for this beam energy, which, 
at the phase transition, 
accompanies a lower freeze-out temperature. 
The picture is consistent
across other centrality classes (not shown), 
which are primarily determined by 
the characteristics of the freeze-out line~\cite{Cleymans:2005xv,STAR:2017sal}.

\begin{figure}[tbp]
\begin{center}
\includegraphics[width=0.9\linewidth]{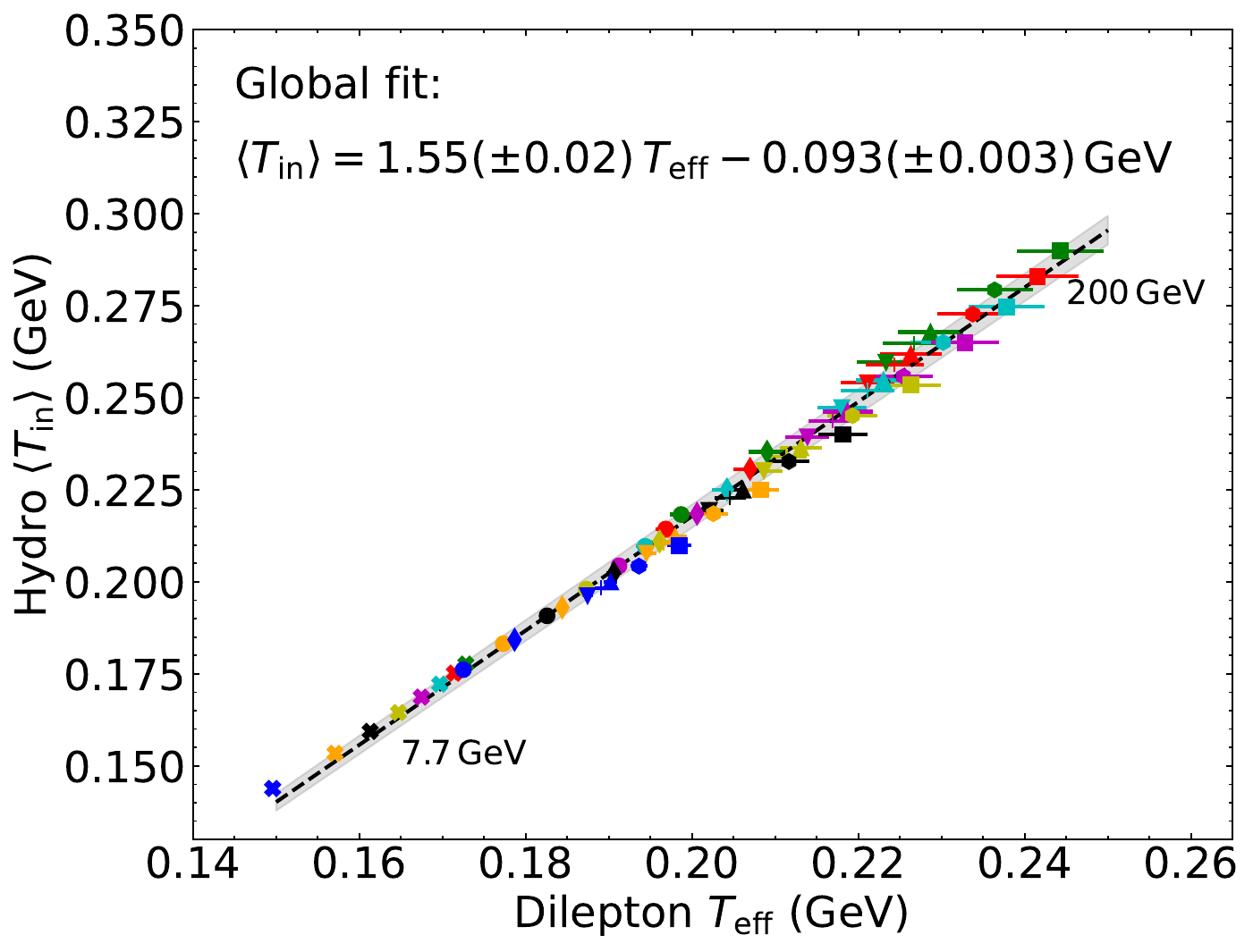}
\caption{%
    Correlation between initial average hydrodynamic temperatures $\langle T_{\rm in}\rangle$ and the derived effective temperature $T_{\rm eff}$ from dilepton spectra for Au+Au collisions at eight beam energies, spanning from 7.7 to 200 GeV, within centrality bins from 0--10\% to 70--80\% (see the Supplemental Material~\cite{suppl} for a table containing all data points). The black dashed line denotes a global fit to all data points,  and the gray band illustrates the uncertainties associated with the fitting procedure. }
    \label{fig:temp_extr}
\end{center}
\end{figure}

While we do not find a significant correlation between \Teff{} and \Tf, 
it appears that \Teff{} is proportional to \Tin. 
Figure~\ref{fig:initial_final_temp} indicates that, 
as the beam energy decreases, both \Teff{} and \Tin{} 
display similar behaviour and
tend to approach each other; again, these trends are consistently observed across various centralities.  
This observation, which was anticipated on theoretical grounds~\cite{Kajantie:1981wg,Hwa:1985xg},
motivates us to investigate the relationship
between the initial mean temperature 
 \Tin{} and the effective temperature \Teff{} in Au+Au collisions at 
a range of beam energies and centralities.
Figure~\ref{fig:temp_extr} demonstrates a 
linear relationship between the initial mean temperature and the effective temperature, 
\begin{equation}\label{eq:linear}
    \langle T_{\rm in}\rangle\ =\ \kappa \, T_{\rm eff} + c \, ,
\end{equation}
where a global fit yields the parameters
\begin{equation*}
     \kappa = 1.55 \pm 0.02 \, , \quad c = - (9.3 \pm 0.3)\times 10^{-2}\ {\rm GeV} \, .
\end{equation*}
This linear relationship presents 
a reliable---and currently unique---means of extracting  the early temperatures of the hot and dense nuclear matter,
utilizing the effective temperature derived from dilepton spectra, 
immune to the distortions caused by Doppler effects. 
Despite the initial temperature variations depicted in Fig.~\ref{fig:initial_final_temp}, 
the definition of mean temperature inherently integrates out these fluctuations, 
and   
the obtained \Tin{} 
is impervious to significant uncertainties via Eq.~\eqref{eq:linear}.
Let us emphasize that \Tin{} should thus be interpreted
as the central value of some broad distribution, whose
statistical variance is relatively large (see Fig.~\ref{fig:temp_extraction_tau} of the supplementary material~\cite{suppl}).
The quoted standard errors, on $\kappa$ and $c\,$,
are due to the fitting procedure alone. 
There exist global studies to address and incorporate systematic model uncertainties \cite{Bernhard:2019bmu,JETSCAPE:2020shq,Nijs:2020ors,Heffernan:2023utr}; future such investigations will incorporate the emission of electromagnetic radiation. 
Importantly, as far as the emission rates are concerned, we 
find negligible variations in \Teff{} from fixing the QCD perturbative coupling.

Several insights can be drawn from the results presented in Fig.~\ref{fig:temp_extr}. 
First and foremost, the unmistakable correlation between \Teff{} and \Tin{},
as opposed to the final mean temperature, 
naturally finds its explanation in the spacetime evolution 
of dilepton production within a rapidly expanding QGP fireball. 
As highlighted in our companion paper~\cite{Churchill:2023vpt}, 
dilepton production is suppressed over time 
because the temperature drops
too fast for the expanding volume to compensate.
Hence, the IMR of dilepton spectra are predominantly 
influenced by the early stages of QGP evolution
and are notably insensitive to the late-stage expansion.

Furthermore, 
uncertainties in \Teff{} tend to grow slightly
with $\sqrt{s_{_{\rm NN}}}$ for
a specific centrality class 
or when transitioning from 
peripheral to central collisions at a given beam energy. 
This trend can be attributed to several factors, 
including the extended lifetime of the QGP at higher 
beam energies or in central collisions, 
combined with more substantial temperature variations 
during its evolution 
(see also Fig.~\ref{fig:temp_extraction_tau} 
in the Supplementary Material \cite{suppl}). 
As a result, 
the dilepton spectra exhibit more pronounced deviations 
from the profiles associated with a specific effective temperature.

Finally, we emphasize that our study primarily focuses on  
thermal dileptons due to the QGP,
whose evolution is described via dissipative hydrodynamics. 
At the lower end of the 
considered beam energies, 
due to the substantial time required for the two colliding nuclei to fully interpenetrate,
the pre-hydrodynamic stage becomes  
non-negligible 
and it is reasonable to assume that its contribution to the dilepton yields increases accordingly. 
In a more suitable dynamical initialization scenario, 
energy continues to be deposited into the collision fireball, 
causing the temperature to increase until 
it reaches a maximum value when the two colliding nuclei have completely traversed each other~\cite{Shen:2017bsr}. 
Subsequently, this is followed by the onset of a pure hydrodynamic QGP expansion. 
Thus, the maximum temperature of 
the pre-hydrodynamic stage and its 
corresponding time should be treated as the initial temperature and starting time for the hydrodynamic evolution in our study. 
Examining the influence of the pre-hydrodynamic stage on dilepton spectra 
and its impact on the associated effective temperature is a topic worthy of dedicated research which we leave for future investigations.

{\it \textbf{Conclusions.}---} 
From the studies reported here, 
using state-of-the-art lepton pair emissivities 
and 
sophisticated (3+1)-dimensional dissipative fluid dynamical modeling, 
it is clear that the 
electromagnetic radiation 
measured in relativistic nuclear collisions 
fulfills its promise of providing penetrating tomographic information of the strongly interacting medium, 
particularly in serving as a thermometer for the early-stage QGP. 
Complementary to the findings in Ref.~\cite{Giacalone:2019ldn} 
that established the general correlation between initial state energy and measured particle multiplicities, 
as well as in Ref.~\cite{Du:2022yok} where the initial baryon density was constrained using rapidity-dependent directed flows, 
our research provides
a solid 
means of probing the  
phase diagram of hot and dense nuclear matter, 
via the slope of dilepton spectra in the IMR and at several beam energies
and centrality classes.
As such, 
our research contributes to filling 
in an essential piece of the puzzle, 
enhancing our understanding
of QCD matter and its manifestation in heavy-ion experiments.

{\it \textbf{Acknowledgements.}---}
We acknowledge very useful conversations with Bailey Forster, Han Gao, and Jean-Fran\c{c}ois Paquet.  This work was funded in part by the U. S. Department of Energy (DOE), under grant No. DE-FG02-00ER41132 (G.~J.), in part by the Agence Nationale de la Recherche,  under grant ANR-22-CE31-0018 (AUTOTHERM) (G.~J.), and in part by the Natural Sciences and Engineering Research Council of Canada (J.~C., L.~D., C.~G., S.~J.). Computations were made on the B\'eluga supercomputer system from McGill University, managed by
Calcul Qu\'ebec and the Digital Research
Alliance of Canada.

\bibliography{ref}

\begin{thebibliography}{80}%
\makeatletter
\providecommand \@ifxundefined [1]{%
 \@ifx{#1\undefined}
}%
\providecommand \@ifnum [1]{%
 \ifnum #1\expandafter \@firstoftwo
 \else \expandafter \@secondoftwo
 \fi
}%
\providecommand \@ifx [1]{%
 \ifx #1\expandafter \@firstoftwo
 \else \expandafter \@secondoftwo
 \fi
}%
\providecommand \natexlab [1]{#1}%
\providecommand \enquote  [1]{``#1''}%
\providecommand \bibnamefont  [1]{#1}%
\providecommand \bibfnamefont [1]{#1}%
\providecommand \citenamefont [1]{#1}%
\providecommand \href@noop [0]{\@secondoftwo}%
\providecommand \href [0]{\begingroup \@sanitize@url \@href}%
\providecommand \@href[1]{\@@startlink{#1}\@@href}%
\providecommand \@@href[1]{\endgroup#1\@@endlink}%
\providecommand \@sanitize@url [0]{\catcode `\\12\catcode `\$12\catcode `\&12\catcode `\#12\catcode `\^12\catcode `\_12\catcode `\%12\relax}%
\providecommand \@@startlink[1]{}%
\providecommand \@@endlink[0]{}%
\providecommand \url  [0]{\begingroup\@sanitize@url \@url }%
\providecommand \@url [1]{\endgroup\@href {#1}{\urlprefix }}%
\providecommand \urlprefix  [0]{URL }%
\providecommand \Eprint [0]{\href }%
\providecommand \doibase [0]{http://dx.doi.org/}%
\providecommand \selectlanguage [0]{\@gobble}%
\providecommand \bibinfo  [0]{\@secondoftwo}%
\providecommand \bibfield  [0]{\@secondoftwo}%
\providecommand \translation [1]{[#1]}%
\providecommand \BibitemOpen [0]{}%
\providecommand \bibitemStop [0]{}%
\providecommand \bibitemNoStop [0]{.\EOS\space}%
\providecommand \EOS [0]{\spacefactor3000\relax}%
\providecommand \BibitemShut  [1]{\csname bibitem#1\endcsname}%
\let\auto@bib@innerbib\@empty
\bibitem [{\citenamefont {Braun-Munzinger}\ and\ \citenamefont {Wambach}(2009)}]{Braun-Munzinger:2008szb}%
  \BibitemOpen
  \bibfield  {author} {\bibinfo {author} {\bibfnamefont {P.}~\bibnamefont {Braun-Munzinger}}\ and\ \bibinfo {author} {\bibfnamefont {J.}~\bibnamefont {Wambach}},\ }\bibfield  {title} {\enquote {\bibinfo {title} {{The Phase Diagram of Strongly-Interacting Matter}},}\ }\href {\doibase 10.1103/RevModPhys.81.1031} {\bibfield  {journal} {\bibinfo  {journal} {Rev. Mod. Phys.}\ }\textbf {\bibinfo {volume} {81}},\ \bibinfo {pages} {1031--1050} (\bibinfo {year} {2009})},\ \Eprint {http://arxiv.org/abs/0801.4256} {arXiv:0801.4256 [hep-ph]} \BibitemShut {NoStop}%
\bibitem [{\citenamefont {Bzdak}\ \emph {et~al.}(2020)\citenamefont {Bzdak}, \citenamefont {Esumi}, \citenamefont {Koch}, \citenamefont {Liao}, \citenamefont {Stephanov},\ and\ \citenamefont {Xu}}]{Bzdak:2019pkr}%
  \BibitemOpen
  \bibfield  {author} {\bibinfo {author} {\bibfnamefont {Adam}\ \bibnamefont {Bzdak}}, \bibinfo {author} {\bibfnamefont {Shinichi}\ \bibnamefont {Esumi}}, \bibinfo {author} {\bibfnamefont {Volker}\ \bibnamefont {Koch}}, \bibinfo {author} {\bibfnamefont {Jinfeng}\ \bibnamefont {Liao}}, \bibinfo {author} {\bibfnamefont {Mikhail}\ \bibnamefont {Stephanov}}, \ and\ \bibinfo {author} {\bibfnamefont {Nu}~\bibnamefont {Xu}},\ }\bibfield  {title} {\enquote {\bibinfo {title} {{Mapping the Phases of Quantum Chromodynamics with Beam Energy Scan}},}\ }\href {\doibase 10.1016/j.physrep.2020.01.005} {\bibfield  {journal} {\bibinfo  {journal} {Phys. Rept.}\ }\textbf {\bibinfo {volume} {853}},\ \bibinfo {pages} {1--87} (\bibinfo {year} {2020})},\ \Eprint {http://arxiv.org/abs/1906.00936} {arXiv:1906.00936 [nucl-th]} \BibitemShut {NoStop}%
\bibitem [{\citenamefont {Shuryak}(2017)}]{Shuryak:2014zxa}%
  \BibitemOpen
  \bibfield  {author} {\bibinfo {author} {\bibfnamefont {Edward}\ \bibnamefont {Shuryak}},\ }\bibfield  {title} {\enquote {\bibinfo {title} {{Strongly coupled quark-gluon plasma in heavy ion collisions}},}\ }\href {\doibase 10.1103/RevModPhys.89.035001} {\bibfield  {journal} {\bibinfo  {journal} {Rev. Mod. Phys.}\ }\textbf {\bibinfo {volume} {89}},\ \bibinfo {pages} {035001} (\bibinfo {year} {2017})},\ \Eprint {http://arxiv.org/abs/1412.8393} {arXiv:1412.8393 [hep-ph]} \BibitemShut {NoStop}%
\bibitem [{\citenamefont {Bernhard}\ \emph {et~al.}(2019)\citenamefont {Bernhard}, \citenamefont {Moreland},\ and\ \citenamefont {Bass}}]{Bernhard:2019bmu}%
  \BibitemOpen
  \bibfield  {author} {\bibinfo {author} {\bibfnamefont {Jonah~E.}\ \bibnamefont {Bernhard}}, \bibinfo {author} {\bibfnamefont {J.~Scott}\ \bibnamefont {Moreland}}, \ and\ \bibinfo {author} {\bibfnamefont {Steffen~A.}\ \bibnamefont {Bass}},\ }\bibfield  {title} {\enquote {\bibinfo {title} {{Bayesian estimation of the specific shear and bulk viscosity of quark\textendash{}gluon plasma}},}\ }\href {\doibase 10.1038/s41567-019-0611-8} {\bibfield  {journal} {\bibinfo  {journal} {Nature Phys.}\ }\textbf {\bibinfo {volume} {15}},\ \bibinfo {pages} {1113--1117} (\bibinfo {year} {2019})}\BibitemShut {NoStop}%
\bibitem [{\citenamefont {Everett}\ \emph {et~al.}(2021)\citenamefont {Everett} \emph {et~al.}}]{JETSCAPE:2020shq}%
  \BibitemOpen
  \bibfield  {author} {\bibinfo {author} {\bibfnamefont {D.}~\bibnamefont {Everett}} \emph {et~al.} (\bibinfo {collaboration} {JETSCAPE}),\ }\bibfield  {title} {\enquote {\bibinfo {title} {{Phenomenological constraints on the transport properties of QCD matter with data-driven model averaging}},}\ }\href {\doibase 10.1103/PhysRevLett.126.242301} {\bibfield  {journal} {\bibinfo  {journal} {Phys. Rev. Lett.}\ }\textbf {\bibinfo {volume} {126}},\ \bibinfo {pages} {242301} (\bibinfo {year} {2021})},\ \Eprint {http://arxiv.org/abs/2010.03928} {arXiv:2010.03928 [hep-ph]} \BibitemShut {NoStop}%
\bibitem [{\citenamefont {Nijs}\ \emph {et~al.}(2021)\citenamefont {Nijs}, \citenamefont {van~der Schee}, \citenamefont {G\"ursoy},\ and\ \citenamefont {Snellings}}]{Nijs:2020ors}%
  \BibitemOpen
  \bibfield  {author} {\bibinfo {author} {\bibfnamefont {Govert}\ \bibnamefont {Nijs}}, \bibinfo {author} {\bibfnamefont {Wilke}\ \bibnamefont {van~der Schee}}, \bibinfo {author} {\bibfnamefont {Umut}\ \bibnamefont {G\"ursoy}}, \ and\ \bibinfo {author} {\bibfnamefont {Raimond}\ \bibnamefont {Snellings}},\ }\bibfield  {title} {\enquote {\bibinfo {title} {{Transverse Momentum Differential Global Analysis of Heavy-Ion Collisions}},}\ }\href {\doibase 10.1103/PhysRevLett.126.202301} {\bibfield  {journal} {\bibinfo  {journal} {Phys. Rev. Lett.}\ }\textbf {\bibinfo {volume} {126}},\ \bibinfo {pages} {202301} (\bibinfo {year} {2021})},\ \Eprint {http://arxiv.org/abs/2010.15130} {arXiv:2010.15130 [nucl-th]} \BibitemShut {NoStop}%
\bibitem [{\citenamefont {Heffernan}\ \emph {et~al.}(2023)\citenamefont {Heffernan}, \citenamefont {Gale}, \citenamefont {Jeon},\ and\ \citenamefont {Paquet}}]{Heffernan:2023utr}%
  \BibitemOpen
  \bibfield  {author} {\bibinfo {author} {\bibfnamefont {Matthew~R.}\ \bibnamefont {Heffernan}}, \bibinfo {author} {\bibfnamefont {Charles}\ \bibnamefont {Gale}}, \bibinfo {author} {\bibfnamefont {Sangyong}\ \bibnamefont {Jeon}}, \ and\ \bibinfo {author} {\bibfnamefont {Jean-Fran\c{c}ois}\ \bibnamefont {Paquet}},\ }\bibfield  {title} {\enquote {\bibinfo {title} {{Bayesian quantification of strongly-interacting matter with color glass condensate initial conditions}},}\ }\href@noop {} {\  (\bibinfo {year} {2023})},\ \Eprint {http://arxiv.org/abs/2302.09478} {arXiv:2302.09478 [nucl-th]} \BibitemShut {NoStop}%
\bibitem [{\citenamefont {Niemi}\ \emph {et~al.}(2016)\citenamefont {Niemi}, \citenamefont {Eskola},\ and\ \citenamefont {Paatelainen}}]{Niemi:2015qia}%
  \BibitemOpen
  \bibfield  {author} {\bibinfo {author} {\bibfnamefont {H.}~\bibnamefont {Niemi}}, \bibinfo {author} {\bibfnamefont {K.~J.}\ \bibnamefont {Eskola}}, \ and\ \bibinfo {author} {\bibfnamefont {R.}~\bibnamefont {Paatelainen}},\ }\bibfield  {title} {\enquote {\bibinfo {title} {{Event-by-event fluctuations in a perturbative QCD + saturation + hydrodynamics model: Determining QCD matter shear viscosity in ultrarelativistic heavy-ion collisions}},}\ }\href {\doibase 10.1103/PhysRevC.93.024907} {\bibfield  {journal} {\bibinfo  {journal} {Phys. Rev. C}\ }\textbf {\bibinfo {volume} {93}},\ \bibinfo {pages} {024907} (\bibinfo {year} {2016})},\ \Eprint {http://arxiv.org/abs/1505.02677} {arXiv:1505.02677 [hep-ph]} \BibitemShut {NoStop}%
\bibitem [{\citenamefont {Hirvonen}\ \emph {et~al.}(2022)\citenamefont {Hirvonen}, \citenamefont {Eskola},\ and\ \citenamefont {Niemi}}]{Hirvonen:2022xfv}%
  \BibitemOpen
  \bibfield  {author} {\bibinfo {author} {\bibfnamefont {H.}~\bibnamefont {Hirvonen}}, \bibinfo {author} {\bibfnamefont {K.~J.}\ \bibnamefont {Eskola}}, \ and\ \bibinfo {author} {\bibfnamefont {H.}~\bibnamefont {Niemi}},\ }\bibfield  {title} {\enquote {\bibinfo {title} {{Flow correlations from a hydrodynamics model with dynamical freeze-out and initial conditions based on perturbative QCD and saturation}},}\ }\href {\doibase 10.1103/PhysRevC.106.044913} {\bibfield  {journal} {\bibinfo  {journal} {Phys. Rev. C}\ }\textbf {\bibinfo {volume} {106}},\ \bibinfo {pages} {044913} (\bibinfo {year} {2022})},\ \Eprint {http://arxiv.org/abs/2206.15207} {arXiv:2206.15207 [hep-ph]} \BibitemShut {NoStop}%
\bibitem [{\citenamefont {Parkkila}\ \emph {et~al.}(2022)\citenamefont {Parkkila}, \citenamefont {Onnerstad}, \citenamefont {Taghavi}, \citenamefont {Mordasini}, \citenamefont {Bilandzic}, \citenamefont {Virta},\ and\ \citenamefont {Kim}}]{Parkkila:2021yha}%
  \BibitemOpen
  \bibfield  {author} {\bibinfo {author} {\bibfnamefont {J.~E.}\ \bibnamefont {Parkkila}}, \bibinfo {author} {\bibfnamefont {A.}~\bibnamefont {Onnerstad}}, \bibinfo {author} {\bibfnamefont {S.~F.}\ \bibnamefont {Taghavi}}, \bibinfo {author} {\bibfnamefont {C.}~\bibnamefont {Mordasini}}, \bibinfo {author} {\bibfnamefont {A.}~\bibnamefont {Bilandzic}}, \bibinfo {author} {\bibfnamefont {M.}~\bibnamefont {Virta}}, \ and\ \bibinfo {author} {\bibfnamefont {D.~J.}\ \bibnamefont {Kim}},\ }\bibfield  {title} {\enquote {\bibinfo {title} {{New constraints for QCD matter from improved Bayesian parameter estimation in heavy-ion collisions at LHC}},}\ }\href {\doibase 10.1016/j.physletb.2022.137485} {\bibfield  {journal} {\bibinfo  {journal} {Phys. Lett. B}\ }\textbf {\bibinfo {volume} {835}},\ \bibinfo {pages} {137485} (\bibinfo {year} {2022})},\ \Eprint {http://arxiv.org/abs/2111.08145} {arXiv:2111.08145 [hep-ph]} \BibitemShut {NoStop}%
\bibitem [{\citenamefont {Parkkila}\ \emph {et~al.}(2021)\citenamefont {Parkkila}, \citenamefont {Onnerstad},\ and\ \citenamefont {Kim}}]{Parkkila:2021tqq}%
  \BibitemOpen
  \bibfield  {author} {\bibinfo {author} {\bibfnamefont {J.~E.}\ \bibnamefont {Parkkila}}, \bibinfo {author} {\bibfnamefont {A.}~\bibnamefont {Onnerstad}}, \ and\ \bibinfo {author} {\bibfnamefont {D.~J.}\ \bibnamefont {Kim}},\ }\bibfield  {title} {\enquote {\bibinfo {title} {{Bayesian estimation of the specific shear and bulk viscosity of the quark-gluon plasma with additional flow harmonic observables}},}\ }\href {\doibase 10.1103/PhysRevC.104.054904} {\bibfield  {journal} {\bibinfo  {journal} {Phys. Rev. C}\ }\textbf {\bibinfo {volume} {104}},\ \bibinfo {pages} {054904} (\bibinfo {year} {2021})},\ \Eprint {http://arxiv.org/abs/2106.05019} {arXiv:2106.05019 [hep-ph]} \BibitemShut {NoStop}%
\bibitem [{\citenamefont {Shuryak}(1978)}]{Shuryak:1978ij}%
  \BibitemOpen
  \bibfield  {author} {\bibinfo {author} {\bibfnamefont {Edward~V.}\ \bibnamefont {Shuryak}},\ }\bibfield  {title} {\enquote {\bibinfo {title} {{Quark-Gluon Plasma and Hadronic Production of Leptons, Photons and Psions}},}\ }\href {\doibase 10.1016/0370-2693(78)90370-2} {\bibfield  {journal} {\bibinfo  {journal} {Phys. Lett. B}\ }\textbf {\bibinfo {volume} {78}},\ \bibinfo {pages} {150} (\bibinfo {year} {1978})}\BibitemShut {NoStop}%
\bibitem [{\citenamefont {Kajantie}\ and\ \citenamefont {Miettinen}(1981)}]{Kajantie:1981wg}%
  \BibitemOpen
  \bibfield  {author} {\bibinfo {author} {\bibfnamefont {K.}~\bibnamefont {Kajantie}}\ and\ \bibinfo {author} {\bibfnamefont {H.~I.}\ \bibnamefont {Miettinen}},\ }\bibfield  {title} {\enquote {\bibinfo {title} {{Temperature Measurement of Quark-Gluon Plasma Formed in High-Energy Nucleus-Nucleus Collisions}},}\ }\href {\doibase 10.1007/BF01548770} {\bibfield  {journal} {\bibinfo  {journal} {Z. Phys. C}\ }\textbf {\bibinfo {volume} {9}},\ \bibinfo {pages} {341} (\bibinfo {year} {1981})}\BibitemShut {NoStop}%
\bibitem [{\citenamefont {McLerran}\ and\ \citenamefont {Toimela}(1985)}]{McLerran1984}%
  \BibitemOpen
  \bibfield  {author} {\bibinfo {author} {\bibfnamefont {Larry~D.}\ \bibnamefont {McLerran}}\ and\ \bibinfo {author} {\bibfnamefont {T.}~\bibnamefont {Toimela}},\ }\bibfield  {title} {\enquote {\bibinfo {title} {{Photon and Dilepton Emission from the Quark - Gluon Plasma: Some General Considerations}},}\ }\href {\doibase 10.1103/PhysRevD.31.545} {\bibfield  {journal} {\bibinfo  {journal} {Phys. Rev. D}\ }\textbf {\bibinfo {volume} {31}},\ \bibinfo {pages} {545} (\bibinfo {year} {1985})}\BibitemShut {NoStop}%
\bibitem [{\citenamefont {Hwa}\ and\ \citenamefont {Kajantie}(1985)}]{Hwa:1985xg}%
  \BibitemOpen
  \bibfield  {author} {\bibinfo {author} {\bibfnamefont {R.~C.}\ \bibnamefont {Hwa}}\ and\ \bibinfo {author} {\bibfnamefont {K.}~\bibnamefont {Kajantie}},\ }\bibfield  {title} {\enquote {\bibinfo {title} {{Diagnosing Quark Matter by Measuring the Total Entropy and the Photon Or Dilepton Emission Rates}},}\ }\href {\doibase 10.1103/PhysRevD.32.1109} {\bibfield  {journal} {\bibinfo  {journal} {Phys. Rev. D}\ }\textbf {\bibinfo {volume} {32}},\ \bibinfo {pages} {1109} (\bibinfo {year} {1985})}\BibitemShut {NoStop}%
\bibitem [{\citenamefont {Kajantie}\ \emph {et~al.}(1986)\citenamefont {Kajantie}, \citenamefont {Kapusta}, \citenamefont {McLerran},\ and\ \citenamefont {Mekjian}}]{Kajantie:1986dh}%
  \BibitemOpen
  \bibfield  {author} {\bibinfo {author} {\bibfnamefont {K.}~\bibnamefont {Kajantie}}, \bibinfo {author} {\bibfnamefont {Joseph~I.}\ \bibnamefont {Kapusta}}, \bibinfo {author} {\bibfnamefont {Larry~D.}\ \bibnamefont {McLerran}}, \ and\ \bibinfo {author} {\bibfnamefont {A.}~\bibnamefont {Mekjian}},\ }\bibfield  {title} {\enquote {\bibinfo {title} {{Dilepton Emission and the QCD Phase Transition in Ultrarelativistic Nuclear Collisions}},}\ }\href {\doibase 10.1103/PhysRevD.34.2746} {\bibfield  {journal} {\bibinfo  {journal} {Phys. Rev. D}\ }\textbf {\bibinfo {volume} {34}},\ \bibinfo {pages} {2746} (\bibinfo {year} {1986})}\BibitemShut {NoStop}%
\bibitem [{\citenamefont {Peitzmann}\ and\ \citenamefont {Thoma}(2002)}]{Peitzmann:2001mz}%
  \BibitemOpen
  \bibfield  {author} {\bibinfo {author} {\bibfnamefont {Thomas}\ \bibnamefont {Peitzmann}}\ and\ \bibinfo {author} {\bibfnamefont {Markus~H.}\ \bibnamefont {Thoma}},\ }\bibfield  {title} {\enquote {\bibinfo {title} {{Direct photons from relativistic heavy ion collisions}},}\ }\href {\doibase 10.1016/S0370-1573(02)00012-1} {\bibfield  {journal} {\bibinfo  {journal} {Phys. Rept.}\ }\textbf {\bibinfo {volume} {364}},\ \bibinfo {pages} {175--246} (\bibinfo {year} {2002})},\ \Eprint {http://arxiv.org/abs/hep-ph/0111114} {arXiv:hep-ph/0111114} \BibitemShut {NoStop}%
\bibitem [{\citenamefont {Salabura}\ and\ \citenamefont {Stroth}(2021)}]{Salabura:2020tou}%
  \BibitemOpen
  \bibfield  {author} {\bibinfo {author} {\bibfnamefont {Piotr}\ \bibnamefont {Salabura}}\ and\ \bibinfo {author} {\bibfnamefont {Joachim}\ \bibnamefont {Stroth}},\ }\bibfield  {title} {\enquote {\bibinfo {title} {{Dilepton radiation from strongly interacting systems}},}\ }\href {\doibase 10.1016/j.ppnp.2021.103869} {\bibfield  {journal} {\bibinfo  {journal} {Prog. Part. Nucl. Phys.}\ }\textbf {\bibinfo {volume} {120}},\ \bibinfo {pages} {103869} (\bibinfo {year} {2021})},\ \Eprint {http://arxiv.org/abs/2005.14589} {arXiv:2005.14589 [nucl-ex]} \BibitemShut {NoStop}%
\bibitem [{\citenamefont {Geurts}\ and\ \citenamefont {Tripolt}(2023)}]{Geurts:2022xmk}%
  \BibitemOpen
  \bibfield  {author} {\bibinfo {author} {\bibfnamefont {Frank}\ \bibnamefont {Geurts}}\ and\ \bibinfo {author} {\bibfnamefont {Ralf-Arno}\ \bibnamefont {Tripolt}},\ }\bibfield  {title} {\enquote {\bibinfo {title} {{Electromagnetic probes: Theory and experiment}},}\ }\href {\doibase 10.1016/j.ppnp.2022.104004} {\bibfield  {journal} {\bibinfo  {journal} {Prog. Part. Nucl. Phys.}\ }\textbf {\bibinfo {volume} {128}},\ \bibinfo {pages} {104004} (\bibinfo {year} {2023})},\ \Eprint {http://arxiv.org/abs/2210.01622} {arXiv:2210.01622 [hep-ph]} \BibitemShut {NoStop}%
\bibitem [{\citenamefont {van Hees}\ \emph {et~al.}(2011)\citenamefont {van Hees}, \citenamefont {Gale},\ and\ \citenamefont {Rapp}}]{vanHees:2011vb}%
  \BibitemOpen
  \bibfield  {author} {\bibinfo {author} {\bibfnamefont {Hendrik}\ \bibnamefont {van Hees}}, \bibinfo {author} {\bibfnamefont {Charles}\ \bibnamefont {Gale}}, \ and\ \bibinfo {author} {\bibfnamefont {Ralf}\ \bibnamefont {Rapp}},\ }\bibfield  {title} {\enquote {\bibinfo {title} {{Thermal Photons and Collective Flow at the Relativistic Heavy-Ion Collider}},}\ }\href {\doibase 10.1103/PhysRevC.84.054906} {\bibfield  {journal} {\bibinfo  {journal} {Phys. Rev. C}\ }\textbf {\bibinfo {volume} {84}},\ \bibinfo {pages} {054906} (\bibinfo {year} {2011})},\ \Eprint {http://arxiv.org/abs/1108.2131} {arXiv:1108.2131 [hep-ph]} \BibitemShut {NoStop}%
\bibitem [{\citenamefont {Shen}\ \emph {et~al.}(2014)\citenamefont {Shen}, \citenamefont {Heinz}, \citenamefont {Paquet},\ and\ \citenamefont {Gale}}]{Shen:2013vja}%
  \BibitemOpen
  \bibfield  {author} {\bibinfo {author} {\bibfnamefont {Chun}\ \bibnamefont {Shen}}, \bibinfo {author} {\bibfnamefont {Ulrich~W}\ \bibnamefont {Heinz}}, \bibinfo {author} {\bibfnamefont {Jean-Francois}\ \bibnamefont {Paquet}}, \ and\ \bibinfo {author} {\bibfnamefont {Charles}\ \bibnamefont {Gale}},\ }\bibfield  {title} {\enquote {\bibinfo {title} {{Thermal photons as a quark-gluon plasma thermometer reexamined}},}\ }\href {\doibase 10.1103/PhysRevC.89.044910} {\bibfield  {journal} {\bibinfo  {journal} {Phys. Rev. C}\ }\textbf {\bibinfo {volume} {89}},\ \bibinfo {pages} {044910} (\bibinfo {year} {2014})},\ \Eprint {http://arxiv.org/abs/1308.2440} {arXiv:1308.2440 [nucl-th]} \BibitemShut {NoStop}%
\bibitem [{Note1()}]{Note1}%
  \BibitemOpen
  \bibinfo {note} {See however the recent study of Ref. \cite {Paquet:2023bdx}.}\BibitemShut {Stop}%
\bibitem [{\citenamefont {Rapp}\ and\ \citenamefont {van Hees}(2016)}]{Rapp:2014hha}%
  \BibitemOpen
  \bibfield  {author} {\bibinfo {author} {\bibfnamefont {Ralf}\ \bibnamefont {Rapp}}\ and\ \bibinfo {author} {\bibfnamefont {Hendrik}\ \bibnamefont {van Hees}},\ }\bibfield  {title} {\enquote {\bibinfo {title} {{Thermal Dileptons as Fireball Thermometer and Chronometer}},}\ }\href {\doibase 10.1016/j.physletb.2015.12.065} {\bibfield  {journal} {\bibinfo  {journal} {Phys. Lett. B}\ }\textbf {\bibinfo {volume} {753}},\ \bibinfo {pages} {586--590} (\bibinfo {year} {2016})},\ \Eprint {http://arxiv.org/abs/1411.4612} {arXiv:1411.4612 [hep-ph]} \BibitemShut {NoStop}%
\bibitem [{\citenamefont {Adamczewski-Musch}\ \emph {et~al.}(2019)\citenamefont {Adamczewski-Musch} \emph {et~al.}}]{HADES:2019auv}%
  \BibitemOpen
  \bibfield  {author} {\bibinfo {author} {\bibfnamefont {J.}~\bibnamefont {Adamczewski-Musch}} \emph {et~al.} (\bibinfo {collaboration} {HADES}),\ }\bibfield  {title} {\enquote {\bibinfo {title} {{Probing dense baryon-rich matter with virtual photons}},}\ }\href {\doibase 10.1038/s41567-019-0583-8} {\bibfield  {journal} {\bibinfo  {journal} {Nature Phys.}\ }\textbf {\bibinfo {volume} {15}},\ \bibinfo {pages} {1040--1045} (\bibinfo {year} {2019})}\BibitemShut {NoStop}%
\bibitem [{Note2()}]{Note2}%
  \BibitemOpen
  \bibinfo {note} {As found in Refs.~\cite {Rapp2013,Vujanovic2013,Vujanovic2016}, the QGP source dominates over dileptons from a hadronic medium for $M > 1.1$~GeV.}\BibitemShut {Stop}%
\bibitem [{\citenamefont {Heinz}\ and\ \citenamefont {Snellings}(2013)}]{Heinz:2013th}%
  \BibitemOpen
  \bibfield  {author} {\bibinfo {author} {\bibfnamefont {Ulrich}\ \bibnamefont {Heinz}}\ and\ \bibinfo {author} {\bibfnamefont {Raimond}\ \bibnamefont {Snellings}},\ }\bibfield  {title} {\enquote {\bibinfo {title} {{Collective flow and viscosity in relativistic heavy-ion collisions}},}\ }\href {\doibase 10.1146/annurev-nucl-102212-170540} {\bibfield  {journal} {\bibinfo  {journal} {Ann. Rev. Nucl. Part. Sci.}\ }\textbf {\bibinfo {volume} {63}},\ \bibinfo {pages} {123--151} (\bibinfo {year} {2013})},\ \Eprint {http://arxiv.org/abs/1301.2826} {arXiv:1301.2826 [nucl-th]} \BibitemShut {NoStop}%
\bibitem [{\citenamefont {Denicol}\ \emph {et~al.}(2018)\citenamefont {Denicol}, \citenamefont {Gale}, \citenamefont {Jeon}, \citenamefont {Monnai}, \citenamefont {Schenke},\ and\ \citenamefont {Shen}}]{Denicol2018}%
  \BibitemOpen
  \bibfield  {author} {\bibinfo {author} {\bibfnamefont {Gabriel~S.}\ \bibnamefont {Denicol}}, \bibinfo {author} {\bibfnamefont {Charles}\ \bibnamefont {Gale}}, \bibinfo {author} {\bibfnamefont {Sangyong}\ \bibnamefont {Jeon}}, \bibinfo {author} {\bibfnamefont {Akihiko}\ \bibnamefont {Monnai}}, \bibinfo {author} {\bibfnamefont {Bj\"orn}\ \bibnamefont {Schenke}}, \ and\ \bibinfo {author} {\bibfnamefont {Chun}\ \bibnamefont {Shen}},\ }\bibfield  {title} {\enquote {\bibinfo {title} {{Net baryon diffusion in fluid dynamic simulations of relativistic heavy-ion collisions}},}\ }\href {\doibase 10.1103/PhysRevC.98.034916} {\bibfield  {journal} {\bibinfo  {journal} {Phys. Rev. C}\ }\textbf {\bibinfo {volume} {98}},\ \bibinfo {pages} {034916} (\bibinfo {year} {2018})},\ \Eprint {http://arxiv.org/abs/1804.10557} {arXiv:1804.10557 [nucl-th]} \BibitemShut {NoStop}%
\bibitem [{\citenamefont {Du}\ and\ \citenamefont {Heinz}(2020)}]{Du:2019obx}%
  \BibitemOpen
  \bibfield  {author} {\bibinfo {author} {\bibfnamefont {Lipei}\ \bibnamefont {Du}}\ and\ \bibinfo {author} {\bibfnamefont {Ulrich}\ \bibnamefont {Heinz}},\ }\bibfield  {title} {\enquote {\bibinfo {title} {{(3+1)-dimensional dissipative relativistic fluid dynamics at non-zero net baryon density}},}\ }\href {\doibase 10.1016/j.cpc.2019.107090} {\bibfield  {journal} {\bibinfo  {journal} {Comput. Phys. Commun.}\ }\textbf {\bibinfo {volume} {251}},\ \bibinfo {pages} {107090} (\bibinfo {year} {2020})},\ \Eprint {http://arxiv.org/abs/1906.11181} {arXiv:1906.11181 [nucl-th]} \BibitemShut {NoStop}%
\bibitem [{\citenamefont {Shen}\ and\ \citenamefont {Alzhrani}(2020)}]{Shen:2020jwv}%
  \BibitemOpen
  \bibfield  {author} {\bibinfo {author} {\bibfnamefont {Chun}\ \bibnamefont {Shen}}\ and\ \bibinfo {author} {\bibfnamefont {Sahr}\ \bibnamefont {Alzhrani}},\ }\bibfield  {title} {\enquote {\bibinfo {title} {{Collision-geometry-based 3D initial condition for relativistic heavy-ion collisions}},}\ }\href {\doibase 10.1103/PhysRevC.102.014909} {\bibfield  {journal} {\bibinfo  {journal} {Phys. Rev. C}\ }\textbf {\bibinfo {volume} {102}},\ \bibinfo {pages} {014909} (\bibinfo {year} {2020})},\ \Eprint {http://arxiv.org/abs/2003.05852} {arXiv:2003.05852 [nucl-th]} \BibitemShut {NoStop}%
\bibitem [{\citenamefont {Monnai}\ \emph {et~al.}(2019)\citenamefont {Monnai}, \citenamefont {Schenke},\ and\ \citenamefont {Shen}}]{Monnai2019}%
  \BibitemOpen
  \bibfield  {author} {\bibinfo {author} {\bibfnamefont {Akihiko}\ \bibnamefont {Monnai}}, \bibinfo {author} {\bibfnamefont {Bj\"orn}\ \bibnamefont {Schenke}}, \ and\ \bibinfo {author} {\bibfnamefont {Chun}\ \bibnamefont {Shen}},\ }\bibfield  {title} {\enquote {\bibinfo {title} {{Equation of state at finite densities for QCD matter in nuclear collisions}},}\ }\href {\doibase 10.1103/PhysRevC.100.024907} {\bibfield  {journal} {\bibinfo  {journal} {Phys. Rev. C}\ }\textbf {\bibinfo {volume} {100}},\ \bibinfo {pages} {024907} (\bibinfo {year} {2019})},\ \Eprint {http://arxiv.org/abs/1902.05095} {arXiv:1902.05095 [nucl-th]} \BibitemShut {NoStop}%
\bibitem [{\citenamefont {Weldon}(1990)}]{Weldon:1990iw}%
  \BibitemOpen
  \bibfield  {author} {\bibinfo {author} {\bibfnamefont {H.~A.}\ \bibnamefont {Weldon}},\ }\bibfield  {title} {\enquote {\bibinfo {title} {{Reformulation of finite temperature dilepton production}},}\ }\href {\doibase 10.1103/PhysRevD.42.2384} {\bibfield  {journal} {\bibinfo  {journal} {Phys. Rev. D}\ }\textbf {\bibinfo {volume} {42}},\ \bibinfo {pages} {2384--2387} (\bibinfo {year} {1990})}\BibitemShut {NoStop}%
\bibitem [{\citenamefont {Gale}\ and\ \citenamefont {Kapusta}(1991)}]{Gale:1990pn}%
  \BibitemOpen
  \bibfield  {author} {\bibinfo {author} {\bibfnamefont {Charles}\ \bibnamefont {Gale}}\ and\ \bibinfo {author} {\bibfnamefont {Joseph~I.}\ \bibnamefont {Kapusta}},\ }\bibfield  {title} {\enquote {\bibinfo {title} {{Vector dominance model at finite temperature}},}\ }\href {\doibase 10.1016/0550-3213(91)90459-B} {\bibfield  {journal} {\bibinfo  {journal} {Nucl. Phys. B}\ }\textbf {\bibinfo {volume} {357}},\ \bibinfo {pages} {65--89} (\bibinfo {year} {1991})}\BibitemShut {NoStop}%
\bibitem [{\citenamefont {Kapusta}\ and\ \citenamefont {Gale}(2011)}]{Kapusta:2006pm}%
  \BibitemOpen
  \bibfield  {author} {\bibinfo {author} {\bibfnamefont {J.~I.}\ \bibnamefont {Kapusta}}\ and\ \bibinfo {author} {\bibfnamefont {Charles}\ \bibnamefont {Gale}},\ }\href {\doibase 10.1017/CBO9780511535130} {\emph {\bibinfo {title} {{Finite-temperature field theory: Principles and applications}}}},\ Cambridge Monographs on Mathematical Physics\ (\bibinfo  {publisher} {Cambridge University Press},\ \bibinfo {year} {2011})\BibitemShut {NoStop}%
\bibitem [{\citenamefont {Laine}\ and\ \citenamefont {Vuorinen}(2016)}]{Laine:2016hma}%
  \BibitemOpen
  \bibfield  {author} {\bibinfo {author} {\bibfnamefont {Mikko}\ \bibnamefont {Laine}}\ and\ \bibinfo {author} {\bibfnamefont {Aleksi}\ \bibnamefont {Vuorinen}},\ }\href {\doibase 10.1007/978-3-319-31933-9} {\emph {\bibinfo {title} {{Basics of Thermal Field Theory}}}},\ Vol.\ \bibinfo {volume} {925}\ (\bibinfo  {publisher} {Springer},\ \bibinfo {year} {2016})\ \Eprint {http://arxiv.org/abs/1701.01554} {arXiv:1701.01554 [hep-ph]} \BibitemShut {NoStop}%
\bibitem [{\citenamefont {Churchill}\ \emph {et~al.}(2023)\citenamefont {Churchill}, \citenamefont {Du}, \citenamefont {Gale}, \citenamefont {Jackson},\ and\ \citenamefont {Jeon}}]{Churchill:2023vpt}%
  \BibitemOpen
  \bibfield  {author} {\bibinfo {author} {\bibfnamefont {Jessica}\ \bibnamefont {Churchill}}, \bibinfo {author} {\bibfnamefont {Lipei}\ \bibnamefont {Du}}, \bibinfo {author} {\bibfnamefont {Charles}\ \bibnamefont {Gale}}, \bibinfo {author} {\bibfnamefont {Greg}\ \bibnamefont {Jackson}}, \ and\ \bibinfo {author} {\bibfnamefont {Sangyong}\ \bibnamefont {Jeon}},\ }\bibfield  {title} {\enquote {\bibinfo {title} {{Dilepton production at NLO and intermediate invariant-mass observables}},}\ }\href@noop {} {\  (\bibinfo {year} {2023})},\ \Eprint {http://arxiv.org/abs/2311.06675} {arXiv:2311.06675 [nucl-th]} \BibitemShut {NoStop}%
\bibitem [{\citenamefont {Ghisoiu}\ and\ \citenamefont {Laine}(2014)}]{Ghisoiu2014}%
  \BibitemOpen
  \bibfield  {author} {\bibinfo {author} {\bibfnamefont {I.}~\bibnamefont {Ghisoiu}}\ and\ \bibinfo {author} {\bibfnamefont {M.}~\bibnamefont {Laine}},\ }\bibfield  {title} {\enquote {\bibinfo {title} {{Interpolation of hard and soft dilepton rates}},}\ }\href {\doibase 10.1007/JHEP10(2014)083} {\bibfield  {journal} {\bibinfo  {journal} {JHEP}\ }\textbf {\bibinfo {volume} {10}},\ \bibinfo {pages} {083} (\bibinfo {year} {2014})},\ \Eprint {http://arxiv.org/abs/1407.7955} {arXiv:1407.7955 [hep-ph]} \BibitemShut {NoStop}%
\bibitem [{Note3()}]{Note3}%
  \BibitemOpen
  \bibinfo {note} {{ The overlap between the regimes cannot be neglected because $\alpha _s$ is not asymptotically small in practice.}}\BibitemShut {Stop}%
\bibitem [{\citenamefont {Braaten}\ and\ \citenamefont {Pisarski}(1990)}]{Braaten:1989mz}%
  \BibitemOpen
  \bibfield  {author} {\bibinfo {author} {\bibfnamefont {Eric}\ \bibnamefont {Braaten}}\ and\ \bibinfo {author} {\bibfnamefont {Robert~D.}\ \bibnamefont {Pisarski}},\ }\bibfield  {title} {\enquote {\bibinfo {title} {{Soft Amplitudes in Hot Gauge Theories: A General Analysis}},}\ }\href {\doibase 10.1016/0550-3213(90)90508-B} {\bibfield  {journal} {\bibinfo  {journal} {Nucl. Phys. B}\ }\textbf {\bibinfo {volume} {337}},\ \bibinfo {pages} {569--634} (\bibinfo {year} {1990})}\BibitemShut {NoStop}%
\bibitem [{\citenamefont {Aurenche}\ \emph {et~al.}(2002{\natexlab{a}})\citenamefont {Aurenche}, \citenamefont {Gelis}, \citenamefont {Moore},\ and\ \citenamefont {Zaraket}}]{Aurenche2002}%
  \BibitemOpen
  \bibfield  {author} {\bibinfo {author} {\bibfnamefont {P.}~\bibnamefont {Aurenche}}, \bibinfo {author} {\bibfnamefont {F.}~\bibnamefont {Gelis}}, \bibinfo {author} {\bibfnamefont {G.~D.}\ \bibnamefont {Moore}}, \ and\ \bibinfo {author} {\bibfnamefont {H.}~\bibnamefont {Zaraket}},\ }\bibfield  {title} {\enquote {\bibinfo {title} {{Landau-Pomeranchuk-Migdal resummation for dilepton production}},}\ }\href {\doibase 10.1088/1126-6708/2002/12/006} {\bibfield  {journal} {\bibinfo  {journal} {JHEP}\ }\textbf {\bibinfo {volume} {12}},\ \bibinfo {pages} {006} (\bibinfo {year} {2002}{\natexlab{a}})},\ \Eprint {http://arxiv.org/abs/hep-ph/0211036} {arXiv:hep-ph/0211036} \BibitemShut {NoStop}%
\bibitem [{\citenamefont {Aurenche}\ \emph {et~al.}(2002{\natexlab{b}})\citenamefont {Aurenche}, \citenamefont {Gelis},\ and\ \citenamefont {Zaraket}}]{Aurenche:2002pc}%
  \BibitemOpen
  \bibfield  {author} {\bibinfo {author} {\bibfnamefont {P.}~\bibnamefont {Aurenche}}, \bibinfo {author} {\bibfnamefont {F.}~\bibnamefont {Gelis}}, \ and\ \bibinfo {author} {\bibfnamefont {H.}~\bibnamefont {Zaraket}},\ }\bibfield  {title} {\enquote {\bibinfo {title} {{Enhanced thermal production of hard dileptons by 3 ---\ensuremath{>} 2 processes}},}\ }\href {\doibase 10.1088/1126-6708/2002/07/063} {\bibfield  {journal} {\bibinfo  {journal} {JHEP}\ }\textbf {\bibinfo {volume} {07}},\ \bibinfo {pages} {063} (\bibinfo {year} {2002}{\natexlab{b}})},\ \Eprint {http://arxiv.org/abs/hep-ph/0204145} {arXiv:hep-ph/0204145} \BibitemShut {NoStop}%
\bibitem [{\citenamefont {Arnold}\ \emph {et~al.}(2001{\natexlab{a}})\citenamefont {Arnold}, \citenamefont {Moore},\ and\ \citenamefont {Yaffe}}]{Arnold2001ba}%
  \BibitemOpen
  \bibfield  {author} {\bibinfo {author} {\bibfnamefont {Peter~Brockway}\ \bibnamefont {Arnold}}, \bibinfo {author} {\bibfnamefont {Guy~D.}\ \bibnamefont {Moore}}, \ and\ \bibinfo {author} {\bibfnamefont {Laurence~G.}\ \bibnamefont {Yaffe}},\ }\bibfield  {title} {\enquote {\bibinfo {title} {{Photon emission from ultrarelativistic plasmas}},}\ }\href {\doibase 10.1088/1126-6708/2001/11/057} {\bibfield  {journal} {\bibinfo  {journal} {JHEP}\ }\textbf {\bibinfo {volume} {11}},\ \bibinfo {pages} {057} (\bibinfo {year} {2001}{\natexlab{a}})},\ \Eprint {http://arxiv.org/abs/hep-ph/0109064} {arXiv:hep-ph/0109064} \BibitemShut {NoStop}%
\bibitem [{\citenamefont {Arnold}\ \emph {et~al.}(2001{\natexlab{b}})\citenamefont {Arnold}, \citenamefont {Moore},\ and\ \citenamefont {Yaffe}}]{Arnold2001ms}%
  \BibitemOpen
  \bibfield  {author} {\bibinfo {author} {\bibfnamefont {Peter~Brockway}\ \bibnamefont {Arnold}}, \bibinfo {author} {\bibfnamefont {Guy~D.}\ \bibnamefont {Moore}}, \ and\ \bibinfo {author} {\bibfnamefont {Laurence~G.}\ \bibnamefont {Yaffe}},\ }\bibfield  {title} {\enquote {\bibinfo {title} {{Photon emission from quark gluon plasma: Complete leading order results}},}\ }\href {\doibase 10.1088/1126-6708/2001/12/009} {\bibfield  {journal} {\bibinfo  {journal} {JHEP}\ }\textbf {\bibinfo {volume} {12}},\ \bibinfo {pages} {009} (\bibinfo {year} {2001}{\natexlab{b}})},\ \Eprint {http://arxiv.org/abs/hep-ph/0111107} {arXiv:hep-ph/0111107} \BibitemShut {NoStop}%
\bibitem [{Note4()}]{Note4}%
  \BibitemOpen
  \bibinfo {note} {To be clear, we are only using the `leading-order' LPM spectral function. Subsequent QCD corrections in this limit have been computed~\cite {Ghiglieri2013,Ghiglieri2014}.}\BibitemShut {Stop}%
\bibitem [{\citenamefont {Baier}\ \emph {et~al.}(1988)\citenamefont {Baier}, \citenamefont {Pire},\ and\ \citenamefont {Schiff}}]{Baier1988}%
  \BibitemOpen
  \bibfield  {author} {\bibinfo {author} {\bibfnamefont {R.}~\bibnamefont {Baier}}, \bibinfo {author} {\bibfnamefont {B.}~\bibnamefont {Pire}}, \ and\ \bibinfo {author} {\bibfnamefont {D.}~\bibnamefont {Schiff}},\ }\bibfield  {title} {\enquote {\bibinfo {title} {{Dilepton production at finite temperature: Perturbative treatment at order $\alpha_s$}},}\ }\href {\doibase 10.1103/PhysRevD.38.2814} {\bibfield  {journal} {\bibinfo  {journal} {Phys. Rev. D}\ }\textbf {\bibinfo {volume} {38}},\ \bibinfo {pages} {2814} (\bibinfo {year} {1988})}\BibitemShut {NoStop}%
\bibitem [{\citenamefont {Gabellini}\ \emph {et~al.}(1990)\citenamefont {Gabellini}, \citenamefont {Grandou},\ and\ \citenamefont {Poizat}}]{Gabellini1989}%
  \BibitemOpen
  \bibfield  {author} {\bibinfo {author} {\bibfnamefont {Y.}~\bibnamefont {Gabellini}}, \bibinfo {author} {\bibfnamefont {T.}~\bibnamefont {Grandou}}, \ and\ \bibinfo {author} {\bibfnamefont {D.}~\bibnamefont {Poizat}},\ }\bibfield  {title} {\enquote {\bibinfo {title} {{Electron - Positron Annihilation in Thermal {QCD}}},}\ }\href {\doibase 10.1016/0003-4916(90)90231-C} {\bibfield  {journal} {\bibinfo  {journal} {Annals Phys.}\ }\textbf {\bibinfo {volume} {202}},\ \bibinfo {pages} {436--466} (\bibinfo {year} {1990})}\BibitemShut {NoStop}%
\bibitem [{\citenamefont {Altherr}\ and\ \citenamefont {Aurenche}(1989)}]{Altherr1989}%
  \BibitemOpen
  \bibfield  {author} {\bibinfo {author} {\bibfnamefont {T.}~\bibnamefont {Altherr}}\ and\ \bibinfo {author} {\bibfnamefont {P.}~\bibnamefont {Aurenche}},\ }\bibfield  {title} {\enquote {\bibinfo {title} {{Finite Temperature QCD Corrections to Lepton Pair Formation in a Quark - Gluon Plasma}},}\ }\href {\doibase 10.1007/BF01556676} {\bibfield  {journal} {\bibinfo  {journal} {Z. Phys. C}\ }\textbf {\bibinfo {volume} {45}},\ \bibinfo {pages} {99} (\bibinfo {year} {1989})}\BibitemShut {NoStop}%
\bibitem [{\citenamefont {Kapusta}\ \emph {et~al.}(1991)\citenamefont {Kapusta}, \citenamefont {Lichard},\ and\ \citenamefont {Seibert}}]{Kapusta1991}%
  \BibitemOpen
  \bibfield  {author} {\bibinfo {author} {\bibfnamefont {Joseph~I.}\ \bibnamefont {Kapusta}}, \bibinfo {author} {\bibfnamefont {P.}~\bibnamefont {Lichard}}, \ and\ \bibinfo {author} {\bibfnamefont {D.}~\bibnamefont {Seibert}},\ }\bibfield  {title} {\enquote {\bibinfo {title} {{High-energy photons from quark - gluon plasma versus hot hadronic gas}},}\ }\href {\doibase 10.1103/PhysRevD.47.4171} {\bibfield  {journal} {\bibinfo  {journal} {Phys. Rev. D}\ }\textbf {\bibinfo {volume} {44}},\ \bibinfo {pages} {2774--2788} (\bibinfo {year} {1991})},\ \bibinfo {note} {[Erratum: Phys.Rev.D 47, 4171 (1993)]}\BibitemShut {NoStop}%
\bibitem [{\citenamefont {Baier}\ \emph {et~al.}(1992)\citenamefont {Baier}, \citenamefont {Nakkagawa}, \citenamefont {Niegawa},\ and\ \citenamefont {Redlich}}]{Baier1991}%
  \BibitemOpen
  \bibfield  {author} {\bibinfo {author} {\bibfnamefont {R.}~\bibnamefont {Baier}}, \bibinfo {author} {\bibfnamefont {H.}~\bibnamefont {Nakkagawa}}, \bibinfo {author} {\bibfnamefont {A.}~\bibnamefont {Niegawa}}, \ and\ \bibinfo {author} {\bibfnamefont {K.}~\bibnamefont {Redlich}},\ }\bibfield  {title} {\enquote {\bibinfo {title} {{Production rate of hard thermal photons and screening of quark mass singularity}},}\ }\href {\doibase 10.1007/BF01625902} {\bibfield  {journal} {\bibinfo  {journal} {Z. Phys. C}\ }\textbf {\bibinfo {volume} {53}},\ \bibinfo {pages} {433--438} (\bibinfo {year} {1992})}\BibitemShut {NoStop}%
\bibitem [{\citenamefont {Laine}(2013{\natexlab{a}})}]{Laine2013vpa}%
  \BibitemOpen
  \bibfield  {author} {\bibinfo {author} {\bibfnamefont {M.}~\bibnamefont {Laine}},\ }\bibfield  {title} {\enquote {\bibinfo {title} {{Thermal 2-loop master spectral function at finite momentum}},}\ }\href {\doibase 10.1007/JHEP05(2013)083} {\bibfield  {journal} {\bibinfo  {journal} {JHEP}\ }\textbf {\bibinfo {volume} {05}},\ \bibinfo {pages} {083} (\bibinfo {year} {2013}{\natexlab{a}})},\ \Eprint {http://arxiv.org/abs/1304.0202} {arXiv:1304.0202 [hep-ph]} \BibitemShut {NoStop}%
\bibitem [{\citenamefont {Laine}(2013{\natexlab{b}})}]{Laine:2013vma}%
  \BibitemOpen
  \bibfield  {author} {\bibinfo {author} {\bibfnamefont {M.}~\bibnamefont {Laine}},\ }\bibfield  {title} {\enquote {\bibinfo {title} {{NLO thermal dilepton rate at non-zero momentum}},}\ }\href {\doibase 10.1007/JHEP11(2013)120} {\bibfield  {journal} {\bibinfo  {journal} {JHEP}\ }\textbf {\bibinfo {volume} {11}},\ \bibinfo {pages} {120} (\bibinfo {year} {2013}{\natexlab{b}})},\ \Eprint {http://arxiv.org/abs/1310.0164} {arXiv:1310.0164 [hep-ph]} \BibitemShut {NoStop}%
\bibitem [{\citenamefont {Jackson}(2019)}]{Jackson2019a}%
  \BibitemOpen
  \bibfield  {author} {\bibinfo {author} {\bibfnamefont {G.}~\bibnamefont {Jackson}},\ }\bibfield  {title} {\enquote {\bibinfo {title} {{Two-loop thermal spectral functions with general kinematics}},}\ }\href {\doibase 10.1103/PhysRevD.100.116019} {\bibfield  {journal} {\bibinfo  {journal} {Phys. Rev. D}\ }\textbf {\bibinfo {volume} {100}},\ \bibinfo {pages} {116019} (\bibinfo {year} {2019})},\ \Eprint {http://arxiv.org/abs/1910.07552} {arXiv:1910.07552 [hep-ph]} \BibitemShut {NoStop}%
\bibitem [{\citenamefont {Jackson}\ and\ \citenamefont {Laine}(2019)}]{Jackson2019}%
  \BibitemOpen
  \bibfield  {author} {\bibinfo {author} {\bibfnamefont {G.}~\bibnamefont {Jackson}}\ and\ \bibinfo {author} {\bibfnamefont {M.}~\bibnamefont {Laine}},\ }\bibfield  {title} {\enquote {\bibinfo {title} {{Testing thermal photon and dilepton rates}},}\ }\href {\doibase 10.1007/JHEP11(2019)144} {\bibfield  {journal} {\bibinfo  {journal} {JHEP}\ }\textbf {\bibinfo {volume} {11}},\ \bibinfo {pages} {144} (\bibinfo {year} {2019})},\ \Eprint {http://arxiv.org/abs/1910.09567} {arXiv:1910.09567 [hep-ph]} \BibitemShut {NoStop}%
\bibitem [{sup()}]{suppl}%
  \BibitemOpen
  \href@noop {} {}\bibinfo {note} {See Supplemental Material at [URL will be inserted by publisher] for a brief description of more technical details of this study}\BibitemShut {NoStop}%
\bibitem [{dil()}]{dileptoncode}%
  \BibitemOpen
  \href@noop {} {}\bibinfo {note} {{DileptonEmission} is a code designed to compute dilepton distributions with next-to-leading-order ({NLO}) emission rates at non-zero chemical potentials, integrated over a hydrodynamic spacetime evolution: {\url{https://github.com/LipeiDu/DileptonEmission}} .}\BibitemShut {Stop}%
\bibitem [{ieb()}]{iebe}%
  \BibitemOpen
  \href@noop {} {}\bibinfo {note} {{iEBE-MUSIC} is a comprehensive and fully-integrated numerical framework designed to streamline hybrid simulations for the study of relativistic heavy-ion collisions: {\url{https://github.com/LipeiDu/iEBE-MUSIC}}.}\BibitemShut {Stop}%
\bibitem [{\citenamefont {Du}\ \emph {et~al.}(2024)\citenamefont {Du}, \citenamefont {Gao}, \citenamefont {Jeon},\ and\ \citenamefont {Gale}}]{Du:2023gnv}%
  \BibitemOpen
  \bibfield  {author} {\bibinfo {author} {\bibfnamefont {Lipei}\ \bibnamefont {Du}}, \bibinfo {author} {\bibfnamefont {Han}\ \bibnamefont {Gao}}, \bibinfo {author} {\bibfnamefont {Sangyong}\ \bibnamefont {Jeon}}, \ and\ \bibinfo {author} {\bibfnamefont {Charles}\ \bibnamefont {Gale}},\ }\bibfield  {title} {\enquote {\bibinfo {title} {{Rapidity scan with multistage hydrodynamic and statistical thermal models}},}\ }\href {\doibase 10.1103/PhysRevC.109.014907} {\bibfield  {journal} {\bibinfo  {journal} {Phys. Rev. C}\ }\textbf {\bibinfo {volume} {109}},\ \bibinfo {pages} {014907} (\bibinfo {year} {2024})},\ \Eprint {http://arxiv.org/abs/2302.13852} {arXiv:2302.13852 [nucl-th]} \BibitemShut {NoStop}%
\bibitem [{Note5()}]{Note5}%
  \BibitemOpen
  \bibinfo {note} {Doing so for the LPM contribution is challenging~\cite {Hauksson:2017udm}, and left for potential future work.}\BibitemShut {Stop}%
\bibitem [{\citenamefont {Adamczyk}\ \emph {et~al.}(2014)\citenamefont {Adamczyk} \emph {et~al.}}]{STAR:2013pwb}%
  \BibitemOpen
  \bibfield  {author} {\bibinfo {author} {\bibfnamefont {L.}~\bibnamefont {Adamczyk}} \emph {et~al.} (\bibinfo {collaboration} {STAR}),\ }\bibfield  {title} {\enquote {\bibinfo {title} {{Dielectron Mass Spectra from Au+Au Collisions at $\sqrt{s_{\rm NN}}$ = 200 GeV}},}\ }\href {\doibase 10.1103/PhysRevLett.113.022301} {\bibfield  {journal} {\bibinfo  {journal} {Phys. Rev. Lett.}\ }\textbf {\bibinfo {volume} {113}},\ \bibinfo {pages} {022301} (\bibinfo {year} {2014})},\ \bibinfo {note} {[Addendum: Phys.Rev.Lett. 113, 049903 (2014)]},\ \Eprint {http://arxiv.org/abs/1312.7397} {arXiv:1312.7397 [hep-ex]} \BibitemShut {NoStop}%
\bibitem [{\citenamefont {Adamczyk}\ \emph {et~al.}(2015{\natexlab{a}})\citenamefont {Adamczyk} \emph {et~al.}}]{STAR2015}%
  \BibitemOpen
  \bibfield  {author} {\bibinfo {author} {\bibfnamefont {L.}~\bibnamefont {Adamczyk}} \emph {et~al.} (\bibinfo {collaboration} {STAR}),\ }\bibfield  {title} {\enquote {\bibinfo {title} {{Measurements of Dielectron Production in Au$+$Au Collisions at $\sqrt{s_{\rm NN}}$ = 200 GeV from the STAR Experiment}},}\ }\href {\doibase 10.1103/PhysRevC.92.024912} {\bibfield  {journal} {\bibinfo  {journal} {Phys. Rev. C}\ }\textbf {\bibinfo {volume} {92}},\ \bibinfo {pages} {024912} (\bibinfo {year} {2015}{\natexlab{a}})},\ \Eprint {http://arxiv.org/abs/1504.01317} {arXiv:1504.01317 [hep-ex]} \BibitemShut {NoStop}%
\bibitem [{\citenamefont {Adamczyk}\ \emph {et~al.}(2015{\natexlab{b}})\citenamefont {Adamczyk} \emph {et~al.}}]{STAR:2015zal}%
  \BibitemOpen
  \bibfield  {author} {\bibinfo {author} {\bibfnamefont {L.}~\bibnamefont {Adamczyk}} \emph {et~al.} (\bibinfo {collaboration} {STAR}),\ }\bibfield  {title} {\enquote {\bibinfo {title} {{Energy dependence of acceptance-corrected dielectron excess mass spectrum at mid-rapidity in Au$+$Au collisions at $\sqrt{s_{NN}} =$ 19.6 and 200 GeV}},}\ }\href {\doibase 10.1016/j.physletb.2015.08.044} {\bibfield  {journal} {\bibinfo  {journal} {Phys. Lett. B}\ }\textbf {\bibinfo {volume} {750}},\ \bibinfo {pages} {64--71} (\bibinfo {year} {2015}{\natexlab{b}})},\ \Eprint {http://arxiv.org/abs/1501.05341} {arXiv:1501.05341 [hep-ex]} \BibitemShut {NoStop}%
\bibitem [{\citenamefont {Abdulhamid}\ \emph {et~al.}(2023)\citenamefont {Abdulhamid} \emph {et~al.}}]{STAR:2023wta}%
  \BibitemOpen
  \bibfield  {author} {\bibinfo {author} {\bibfnamefont {M.~I.}\ \bibnamefont {Abdulhamid}} \emph {et~al.} (\bibinfo {collaboration} {STAR}),\ }\bibfield  {title} {\enquote {\bibinfo {title} {{Measurements of dielectron production in Au+Au collisions at sNN=27, 39, and 62.4 GeV from the STAR experiment}},}\ }\href {\doibase 10.1103/PhysRevC.107.L061901} {\bibfield  {journal} {\bibinfo  {journal} {Phys. Rev. C}\ }\textbf {\bibinfo {volume} {107}},\ \bibinfo {pages} {L061901} (\bibinfo {year} {2023})}\BibitemShut {NoStop}%
\bibitem [{\citenamefont {Cleymans}\ \emph {et~al.}(2006)\citenamefont {Cleymans}, \citenamefont {Oeschler}, \citenamefont {Redlich},\ and\ \citenamefont {Wheaton}}]{Cleymans:2005xv}%
  \BibitemOpen
  \bibfield  {author} {\bibinfo {author} {\bibfnamefont {J.}~\bibnamefont {Cleymans}}, \bibinfo {author} {\bibfnamefont {H.}~\bibnamefont {Oeschler}}, \bibinfo {author} {\bibfnamefont {K.}~\bibnamefont {Redlich}}, \ and\ \bibinfo {author} {\bibfnamefont {S.}~\bibnamefont {Wheaton}},\ }\bibfield  {title} {\enquote {\bibinfo {title} {{Comparison of chemical freeze-out criteria in heavy-ion collisions}},}\ }\href {\doibase 10.1103/PhysRevC.73.034905} {\bibfield  {journal} {\bibinfo  {journal} {Phys. Rev. C}\ }\textbf {\bibinfo {volume} {73}},\ \bibinfo {pages} {034905} (\bibinfo {year} {2006})},\ \Eprint {http://arxiv.org/abs/hep-ph/0511094} {arXiv:hep-ph/0511094} \BibitemShut {NoStop}%
\bibitem [{\citenamefont {Adamczyk}\ \emph {et~al.}(2017)\citenamefont {Adamczyk} \emph {et~al.}}]{STAR:2017sal}%
  \BibitemOpen
  \bibfield  {author} {\bibinfo {author} {\bibfnamefont {L.}~\bibnamefont {Adamczyk}} \emph {et~al.} (\bibinfo {collaboration} {STAR}),\ }\bibfield  {title} {\enquote {\bibinfo {title} {{Bulk Properties of the Medium Produced in Relativistic Heavy-Ion Collisions from the Beam Energy Scan Program}},}\ }\href {\doibase 10.1103/PhysRevC.96.044904} {\bibfield  {journal} {\bibinfo  {journal} {Phys. Rev. C}\ }\textbf {\bibinfo {volume} {96}},\ \bibinfo {pages} {044904} (\bibinfo {year} {2017})},\ \Eprint {http://arxiv.org/abs/1701.07065} {arXiv:1701.07065 [nucl-ex]} \BibitemShut {NoStop}%
\bibitem [{Note6()}]{Note6}%
  \BibitemOpen
  \bibinfo {note} {Those are labeled ``cocktail'' by the experimental collaboration. They are lepton pairs coming from the Drell-Yan process, from semi-leptonic decays of open flavor mesons, and from radiative decays of final-state hadrons.}\BibitemShut {Stop}%
\bibitem [{\citenamefont {Rapp}\ \emph {et~al.}(2010)\citenamefont {Rapp}, \citenamefont {Wambach},\ and\ \citenamefont {van Hees}}]{Rapp:2009yu}%
  \BibitemOpen
  \bibfield  {author} {\bibinfo {author} {\bibfnamefont {R.}~\bibnamefont {Rapp}}, \bibinfo {author} {\bibfnamefont {J.}~\bibnamefont {Wambach}}, \ and\ \bibinfo {author} {\bibfnamefont {H.}~\bibnamefont {van Hees}},\ }\bibfield  {title} {\enquote {\bibinfo {title} {{The Chiral Restoration Transition of QCD and Low Mass Dileptons}},}\ }\href {\doibase 10.1007/978-3-642-01539-7_6} {\bibfield  {journal} {\bibinfo  {journal} {Landolt-Bornstein}\ }\textbf {\bibinfo {volume} {23}},\ \bibinfo {pages} {134} (\bibinfo {year} {2010})},\ \Eprint {http://arxiv.org/abs/0901.3289} {arXiv:0901.3289 [hep-ph]} \BibitemShut {NoStop}%
\bibitem [{\citenamefont {Shen}\ and\ \citenamefont {Schenke}(2018)}]{Shen:2017bsr}%
  \BibitemOpen
  \bibfield  {author} {\bibinfo {author} {\bibfnamefont {Chun}\ \bibnamefont {Shen}}\ and\ \bibinfo {author} {\bibfnamefont {Bj\"orn}\ \bibnamefont {Schenke}},\ }\bibfield  {title} {\enquote {\bibinfo {title} {{Dynamical initial state model for relativistic heavy-ion collisions}},}\ }\href {\doibase 10.1103/PhysRevC.97.024907} {\bibfield  {journal} {\bibinfo  {journal} {Phys. Rev. C}\ }\textbf {\bibinfo {volume} {97}},\ \bibinfo {pages} {024907} (\bibinfo {year} {2018})},\ \Eprint {http://arxiv.org/abs/1710.00881} {arXiv:1710.00881 [nucl-th]} \BibitemShut {NoStop}%
\bibitem [{\citenamefont {Giacalone}\ \emph {et~al.}(2019)\citenamefont {Giacalone}, \citenamefont {Mazeliauskas},\ and\ \citenamefont {Schlichting}}]{Giacalone:2019ldn}%
  \BibitemOpen
  \bibfield  {author} {\bibinfo {author} {\bibfnamefont {Giuliano}\ \bibnamefont {Giacalone}}, \bibinfo {author} {\bibfnamefont {Aleksas}\ \bibnamefont {Mazeliauskas}}, \ and\ \bibinfo {author} {\bibfnamefont {S\"oren}\ \bibnamefont {Schlichting}},\ }\bibfield  {title} {\enquote {\bibinfo {title} {{Hydrodynamic attractors, initial state energy and particle production in relativistic nuclear collisions}},}\ }\href {\doibase 10.1103/PhysRevLett.123.262301} {\bibfield  {journal} {\bibinfo  {journal} {Phys. Rev. Lett.}\ }\textbf {\bibinfo {volume} {123}},\ \bibinfo {pages} {262301} (\bibinfo {year} {2019})},\ \Eprint {http://arxiv.org/abs/1908.02866} {arXiv:1908.02866 [hep-ph]} \BibitemShut {NoStop}%
\bibitem [{\citenamefont {Du}\ \emph {et~al.}(2023)\citenamefont {Du}, \citenamefont {Shen}, \citenamefont {Jeon},\ and\ \citenamefont {Gale}}]{Du:2022yok}%
  \BibitemOpen
  \bibfield  {author} {\bibinfo {author} {\bibfnamefont {Lipei}\ \bibnamefont {Du}}, \bibinfo {author} {\bibfnamefont {Chun}\ \bibnamefont {Shen}}, \bibinfo {author} {\bibfnamefont {Sangyong}\ \bibnamefont {Jeon}}, \ and\ \bibinfo {author} {\bibfnamefont {Charles}\ \bibnamefont {Gale}},\ }\bibfield  {title} {\enquote {\bibinfo {title} {{Probing initial baryon stopping and equation~of state with rapidity-dependent directed flow of identified particles}},}\ }\href {\doibase 10.1103/PhysRevC.108.L041901} {\bibfield  {journal} {\bibinfo  {journal} {Phys. Rev. C}\ }\textbf {\bibinfo {volume} {108}},\ \bibinfo {pages} {L041901} (\bibinfo {year} {2023})},\ \Eprint {http://arxiv.org/abs/2211.16408} {arXiv:2211.16408 [nucl-th]} \BibitemShut {NoStop}%
\bibitem [{\citenamefont {Paquet}(2023)}]{Paquet:2023bdx}%
  \BibitemOpen
  \bibfield  {author} {\bibinfo {author} {\bibfnamefont {Jean-Fran\c{c}ois}\ \bibnamefont {Paquet}},\ }\bibfield  {title} {\enquote {\bibinfo {title} {{Thermal photon production in Gubser inviscid relativistic fluid dynamics}},}\ }\href@noop {} {\  (\bibinfo {year} {2023})},\ \Eprint {http://arxiv.org/abs/2305.10669} {arXiv:2305.10669 [nucl-th]} \BibitemShut {NoStop}%
\bibitem [{\citenamefont {Rapp}(2013)}]{Rapp2013}%
  \BibitemOpen
  \bibfield  {author} {\bibinfo {author} {\bibfnamefont {Ralf}\ \bibnamefont {Rapp}},\ }\bibfield  {title} {\enquote {\bibinfo {title} {{Dilepton Spectroscopy of QCD Matter at Collider Energies}},}\ }\href {\doibase 10.1155/2013/148253} {\bibfield  {journal} {\bibinfo  {journal} {Adv. High Energy Phys.}\ }\textbf {\bibinfo {volume} {2013}},\ \bibinfo {pages} {148253} (\bibinfo {year} {2013})},\ \Eprint {http://arxiv.org/abs/1304.2309} {arXiv:1304.2309 [hep-ph]} \BibitemShut {NoStop}%
\bibitem [{\citenamefont {Vujanovic}\ \emph {et~al.}(2014)\citenamefont {Vujanovic}, \citenamefont {Young}, \citenamefont {Schenke}, \citenamefont {Rapp}, \citenamefont {Jeon},\ and\ \citenamefont {Gale}}]{Vujanovic2013}%
  \BibitemOpen
  \bibfield  {author} {\bibinfo {author} {\bibfnamefont {Gojko}\ \bibnamefont {Vujanovic}}, \bibinfo {author} {\bibfnamefont {Clint}\ \bibnamefont {Young}}, \bibinfo {author} {\bibfnamefont {Bjoern}\ \bibnamefont {Schenke}}, \bibinfo {author} {\bibfnamefont {Ralf}\ \bibnamefont {Rapp}}, \bibinfo {author} {\bibfnamefont {Sangyong}\ \bibnamefont {Jeon}}, \ and\ \bibinfo {author} {\bibfnamefont {Charles}\ \bibnamefont {Gale}},\ }\bibfield  {title} {\enquote {\bibinfo {title} {{Dilepton emission in high-energy heavy-ion collisions with viscous hydrodynamics}},}\ }\href {\doibase 10.1103/PhysRevC.89.034904} {\bibfield  {journal} {\bibinfo  {journal} {Phys. Rev. C}\ }\textbf {\bibinfo {volume} {89}},\ \bibinfo {pages} {034904} (\bibinfo {year} {2014})},\ \Eprint {http://arxiv.org/abs/1312.0676} {arXiv:1312.0676 [nucl-th]} \BibitemShut {NoStop}%
\bibitem [{\citenamefont {Vujanovic}\ \emph {et~al.}(2016)\citenamefont {Vujanovic}, \citenamefont {Paquet}, \citenamefont {Denicol}, \citenamefont {Luzum}, \citenamefont {Jeon},\ and\ \citenamefont {Gale}}]{Vujanovic2016}%
  \BibitemOpen
  \bibfield  {author} {\bibinfo {author} {\bibfnamefont {Gojko}\ \bibnamefont {Vujanovic}}, \bibinfo {author} {\bibfnamefont {Jean-Fran\c{c}ois}\ \bibnamefont {Paquet}}, \bibinfo {author} {\bibfnamefont {Gabriel~S.}\ \bibnamefont {Denicol}}, \bibinfo {author} {\bibfnamefont {Matthew}\ \bibnamefont {Luzum}}, \bibinfo {author} {\bibfnamefont {Sangyong}\ \bibnamefont {Jeon}}, \ and\ \bibinfo {author} {\bibfnamefont {Charles}\ \bibnamefont {Gale}},\ }\bibfield  {title} {\enquote {\bibinfo {title} {{Electromagnetic radiation as a probe of the initial state and of viscous dynamics in relativistic nuclear collisions}},}\ }\href {\doibase 10.1103/PhysRevC.94.014904} {\bibfield  {journal} {\bibinfo  {journal} {Phys. Rev. C}\ }\textbf {\bibinfo {volume} {94}},\ \bibinfo {pages} {014904} (\bibinfo {year} {2016})},\ \Eprint {http://arxiv.org/abs/1602.01455} {arXiv:1602.01455 [nucl-th]} \BibitemShut {NoStop}%
\bibitem [{\citenamefont {Ghiglieri}\ \emph {et~al.}(2013)\citenamefont {Ghiglieri}, \citenamefont {Hong}, \citenamefont {Kurkela}, \citenamefont {Lu}, \citenamefont {Moore},\ and\ \citenamefont {Teaney}}]{Ghiglieri2013}%
  \BibitemOpen
  \bibfield  {author} {\bibinfo {author} {\bibfnamefont {Jacopo}\ \bibnamefont {Ghiglieri}}, \bibinfo {author} {\bibfnamefont {Juhee}\ \bibnamefont {Hong}}, \bibinfo {author} {\bibfnamefont {Aleksi}\ \bibnamefont {Kurkela}}, \bibinfo {author} {\bibfnamefont {Egang}\ \bibnamefont {Lu}}, \bibinfo {author} {\bibfnamefont {Guy~D.}\ \bibnamefont {Moore}}, \ and\ \bibinfo {author} {\bibfnamefont {Derek}\ \bibnamefont {Teaney}},\ }\bibfield  {title} {\enquote {\bibinfo {title} {{Next-to-leading order thermal photon production in a weakly coupled quark-gluon plasma}},}\ }\href {\doibase 10.1007/JHEP05(2013)010} {\bibfield  {journal} {\bibinfo  {journal} {JHEP}\ }\textbf {\bibinfo {volume} {05}},\ \bibinfo {pages} {010} (\bibinfo {year} {2013})},\ \Eprint {http://arxiv.org/abs/1302.5970} {arXiv:1302.5970 [hep-ph]} \BibitemShut {NoStop}%
\bibitem [{\citenamefont {Ghiglieri}\ and\ \citenamefont {Moore}(2014)}]{Ghiglieri2014}%
  \BibitemOpen
  \bibfield  {author} {\bibinfo {author} {\bibfnamefont {Jacopo}\ \bibnamefont {Ghiglieri}}\ and\ \bibinfo {author} {\bibfnamefont {Guy~D.}\ \bibnamefont {Moore}},\ }\bibfield  {title} {\enquote {\bibinfo {title} {{Low Mass Thermal Dilepton Production at NLO in a Weakly Coupled Quark-Gluon Plasma}},}\ }\href {\doibase 10.1007/JHEP12(2014)029} {\bibfield  {journal} {\bibinfo  {journal} {JHEP}\ }\textbf {\bibinfo {volume} {12}},\ \bibinfo {pages} {029} (\bibinfo {year} {2014})},\ \Eprint {http://arxiv.org/abs/1410.4203} {arXiv:1410.4203 [hep-ph]} \BibitemShut {NoStop}%
\bibitem [{\citenamefont {Hauksson}\ \emph {et~al.}(2018)\citenamefont {Hauksson}, \citenamefont {Jeon},\ and\ \citenamefont {Gale}}]{Hauksson:2017udm}%
  \BibitemOpen
  \bibfield  {author} {\bibinfo {author} {\bibfnamefont {Sigtryggur}\ \bibnamefont {Hauksson}}, \bibinfo {author} {\bibfnamefont {Sangyong}\ \bibnamefont {Jeon}}, \ and\ \bibinfo {author} {\bibfnamefont {Charles}\ \bibnamefont {Gale}},\ }\bibfield  {title} {\enquote {\bibinfo {title} {{Photon emission from quark-gluon plasma out of equilibrium}},}\ }\href {\doibase 10.1103/PhysRevC.97.014901} {\bibfield  {journal} {\bibinfo  {journal} {Phys. Rev. C}\ }\textbf {\bibinfo {volume} {97}},\ \bibinfo {pages} {014901} (\bibinfo {year} {2018})},\ \Eprint {http://arxiv.org/abs/1709.03598} {arXiv:1709.03598 [nucl-th]} \BibitemShut {NoStop}%
\bibitem [{\citenamefont {Baikov}\ \emph {et~al.}(2017)\citenamefont {Baikov}, \citenamefont {Chetyrkin},\ and\ \citenamefont {K\"uhn}}]{Baikov2016}%
  \BibitemOpen
  \bibfield  {author} {\bibinfo {author} {\bibfnamefont {P.~A.}\ \bibnamefont {Baikov}}, \bibinfo {author} {\bibfnamefont {K.~G.}\ \bibnamefont {Chetyrkin}}, \ and\ \bibinfo {author} {\bibfnamefont {J.~H.}\ \bibnamefont {K\"uhn}},\ }\bibfield  {title} {\enquote {\bibinfo {title} {{Five-Loop Running of the QCD coupling constant}},}\ }\href {\doibase 10.1103/PhysRevLett.118.082002} {\bibfield  {journal} {\bibinfo  {journal} {Phys. Rev. Lett.}\ }\textbf {\bibinfo {volume} {118}},\ \bibinfo {pages} {082002} (\bibinfo {year} {2017})},\ \Eprint {http://arxiv.org/abs/1606.08659} {arXiv:1606.08659 [hep-ph]} \BibitemShut {NoStop}%
\bibitem [{\citenamefont {Aoki}\ \emph {et~al.}(2020)\citenamefont {Aoki} \emph {et~al.}}]{Aoki2019}%
  \BibitemOpen
  \bibfield  {author} {\bibinfo {author} {\bibfnamefont {S.}~\bibnamefont {Aoki}} \emph {et~al.} (\bibinfo {collaboration} {Flavour Lattice Averaging Group}),\ }\bibfield  {title} {\enquote {\bibinfo {title} {{FLAG Review 2019}},}\ }\href {\doibase 10.1140/epjc/s10052-019-7354-7} {\bibfield  {journal} {\bibinfo  {journal} {Eur. Phys. J.}\ }\textbf {\bibinfo {volume} {C80}},\ \bibinfo {pages} {113} (\bibinfo {year} {2020})},\ \Eprint {http://arxiv.org/abs/1902.08191} {arXiv:1902.08191 [hep-lat]} \BibitemShut {NoStop}%
\bibitem [{\citenamefont {Jackson}(2022)}]{Jackson2022}%
  \BibitemOpen
  \bibfield  {author} {\bibinfo {author} {\bibfnamefont {Greg}\ \bibnamefont {Jackson}},\ }\bibfield  {title} {\enquote {\bibinfo {title} {{Shedding light on thermal photon and dilepton production}},}\ }\href {\doibase 10.1051/epjconf/202227405014} {\bibfield  {journal} {\bibinfo  {journal} {EPJ Web Conf.}\ }\textbf {\bibinfo {volume} {274}},\ \bibinfo {pages} {05014} (\bibinfo {year} {2022})},\ \Eprint {http://arxiv.org/abs/2211.09575} {arXiv:2211.09575 [hep-ph]} \BibitemShut {NoStop}%
\bibitem [{\citenamefont {Bala}\ \emph {et~al.}(2023)\citenamefont {Bala}, \citenamefont {Ali}, \citenamefont {Francis}, \citenamefont {Jackson}, \citenamefont {Kaczmarek},\ and\ \citenamefont {Ueding}}]{Bala2022}%
  \BibitemOpen
  \bibfield  {author} {\bibinfo {author} {\bibfnamefont {Dibyendu}\ \bibnamefont {Bala}}, \bibinfo {author} {\bibfnamefont {Sajid}\ \bibnamefont {Ali}}, \bibinfo {author} {\bibfnamefont {Anthony}\ \bibnamefont {Francis}}, \bibinfo {author} {\bibfnamefont {Greg}\ \bibnamefont {Jackson}}, \bibinfo {author} {\bibfnamefont {Olaf}\ \bibnamefont {Kaczmarek}}, \ and\ \bibinfo {author} {\bibfnamefont {Tristan}\ \bibnamefont {Ueding}},\ }\bibfield  {title} {\enquote {\bibinfo {title} {{Photon production rate from Transverse-Longitudinal (T$-$L) mesonic correlator on the lattice}},}\ }\href {\doibase 10.22323/1.430.0169} {\bibfield  {journal} {\bibinfo  {journal} {PoS}\ }\textbf {\bibinfo {volume} {LATTICE2022}},\ \bibinfo {pages} {169} (\bibinfo {year} {2023})},\ \Eprint {http://arxiv.org/abs/2212.11509} {arXiv:2212.11509 [hep-lat]} \BibitemShut {NoStop}%
\bibitem [{\citenamefont {Du}(2023)}]{Du:2023efk}%
  \BibitemOpen
  \bibfield  {author} {\bibinfo {author} {\bibfnamefont {Lipei}\ \bibnamefont {Du}},\ }\bibfield  {title} {\enquote {\bibinfo {title} {{Bulk medium properties of heavy-ion collisions at beam energy scan with a multistage hydrodynamic model}},}\ }\href@noop {} {\  (\bibinfo {year} {2023})},\ \Eprint {http://arxiv.org/abs/2401.00596} {arXiv:2401.00596 [hep-ph]} \BibitemShut {NoStop}%
\end{thebibliography}%


\clearpage
\appendix*

\onecolumngrid
\renewcommand{\thefigure}{S\arabic{figure}}
\setcounter{figure}{0} 
\setcounter{equation}{0} 
\renewcommand{\thetable}{S\arabic{table}}


\section*{Supplemental Material}

\subsection{Running coupling\label{sec:running}}

\newcommand{\tinymsbar}{{\overline{\mbox{\tiny\rm{MS}}}}}
\newcommand{\Lambdamsbar}{{\Lambda_\tinymsbar}}

\begin{figure*}[!htbp]
\begin{center}
\includegraphics[width= .49\linewidth]{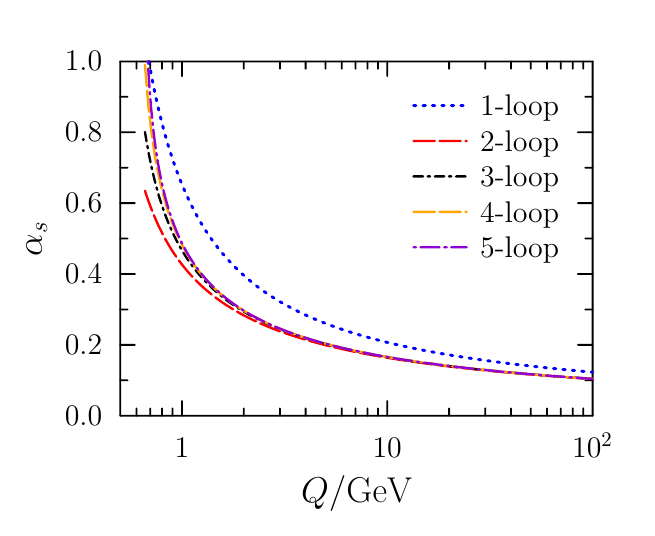}
\includegraphics[width= .49\linewidth]{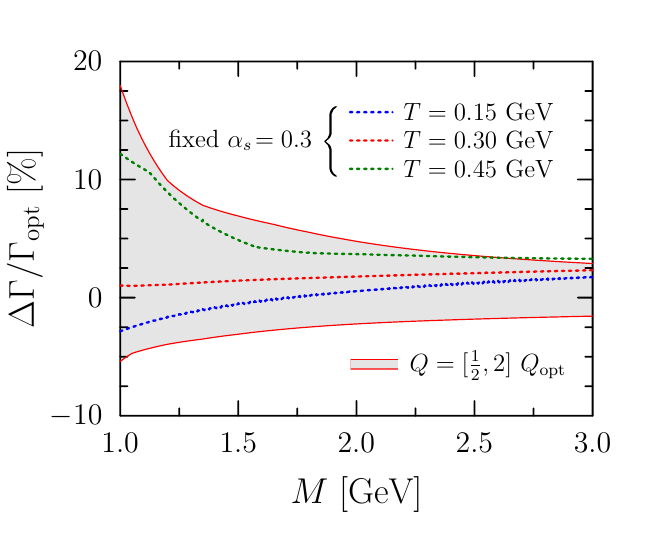}
\caption{%
       Left: Value of the QCD coupling as a function of the renormalisation scale $Q\,$, at various loop levels.
       Right: Illustration of the percentual effect from the running coupling, in the IMR.
       Here we show a ratio, taken w.r.t. to the observable $\dd \Gamma/\dd M$
       for a point source (cf. Fig.~6 in Ref.~\cite{Churchill:2023vpt}).
       The denominator is defined by using the 5-loop running coupling, evaluated at $Q=Q_{\rm opt}$ from \eqref{scale},
       and the gray band shows the deviation from varying this choice by a factor of 2 (for $T=0.30$~GeV).
       Using a fixed coupling $\alpha_s=0.3$ gives the 
       dotted lines, shown for $T=0.15,\,0.30,\,0.45$~GeV.
       (In this figure, $\muB=0$.)
    }
    \label{fig:rate_with_running}
\end{center}
\end{figure*}

For phenomenology,
the value of the QCD gauge coupling
in our approach
needs to be specified. 
The coupling is a function of the energy scale $Q$
and obeys the renormalization group equation
 $\pd_t a_s = - \sum_{k=1}^\ell \beta_{k-1} a_s^{k+1}$
to $\ell$-loop accuracy,
where $a_s \equiv \alpha_s(Q)/\pi$ 
and $t \equiv \log (Q^2 / \Lambdamsbar^{\!\!\!\!\!\!\!2}\ \,)\,$.
The coefficients $\beta_0,\beta_1,\beta_2,\beta_3, \beta_4$ can be found in Ref.~\cite{Baikov2016}.
Figure~\ref{fig:rate_with_running} (left) 
shows the resulting
dependence on $Q$ in GeV, for $\ell \leq 5\,$.
(For $\Nf=3$ we use $\Lambdamsbar = 343$ MeV to set physical units~\cite{Aoki2019}.)
In the calculation of the dilepton rate,
we should evaluate $\alpha_s(Q)$ at the `optimal' scale
which should depend on those among $M,\,T,\,\muB\,$.
Rigorously, the choice of $Q$ should emerge from 
higher order QCD corrections than we are considering.
Since we make use of the spectral function in the IMR,
we have to be pragmatic and adopt a procedure 
which is compatible with the limits in which $Q$ is known.
For example, one could take the geometric mean
(a slightly different choice was made in Ref.~\cite{Ghisoiu2014}):
\begin{eqnarray}
  Q_{\rm opt} &=&
  \sqrt{ M^2 + (2\pi T)^2 + \tfrac19\, {\muB^2} }  \ .
\label{scale}
\end{eqnarray}
This choice is motivated by the fact that 
when one of the three scales $M\,$, $T$ or $\muB$
is much larger than the other two, then $Q_{\rm opt}$ should
be set by that dominant parameter.
At $\muB=0$, this choice was also found to give
relatively good agreement with non-perturbative lattice data
for the Euclidean correlator~\cite{Jackson2019,Jackson2022,Bala2022}.

In order to assess the uncertainty in our perturbative predictions,
we vary the renormalization scale in the range
$Q\in [\frac12, 2] \times Q_{\rm opt}\,$.
To avoid intrinsic difficulties of QCD in the far infrared,
which reflects in \eqref{scale} when $M\,$, $T$ and $\muB$ are simultaneously small, 
we do not allow the coupling to exceed unity: $\alpha_s \leq 1\,$.
(The corresponding dilepton rates in this regime are in any case small, and
do not contribute significantly to the total yield.)
For simplicity, we consider the $\bm k$-integrated rate
from a static point source as done in Ref.~\cite{Churchill:2023vpt}.
In this setting, we obtain the right hand side plot in Fig.~\ref{fig:rate_with_running}
which depicts the relative importance
of the running coupling as a function of $M\,$.
We can make two observations:
Firstly, the sensitivity to varying the scale (at $T=0.3$~GeV) diminishes
from $10\%$ for $M\lsim 1$~GeV to $2\%$ for $M\lsim 3$~GeV.
Secondly, using a fixed value of $\alpha_s =0.3$ appears 
to be accurate to a few percent for $M\gsim 1.5$~GeV.

\subsection{Temperature extraction in time}

In this section, we elaborate on the methodology for conducting a comparative analysis between \Teff{} derived from dilepton spectra and the evolving hydrodynamic temperatures at individual time steps. For \Teff{}, we compute dilepton production from all fluid cells at each time step and extract the corresponding \Teff{} values from the spectra. To quantify the spatially fluctuating hydrodynamic temperatures, which arise from the system's intrinsic inhomogeneity, we compute both the mean (\Tavg{}) and standard deviation ($\langle \delta T\rangle$) for fluid cells at each time step. In these calculations, the temperature of each fluid cell (within spacetime rapidity window $|\eta_s|<1$) is weighted by its energy density ($e$) multiplied by the Lorentz boost factor ($\gamma=u^t$), following what is done in the MUSIC package. The weighting factor considers that fluid cells with varying temperatures do not contribute equally to the observables. Instead, their contribution is proportional to their energy (or entropy) density.

\begin{figure*}[!htbp]
\begin{center}
\vspace{.5cm}
\includegraphics[width= .45\linewidth]{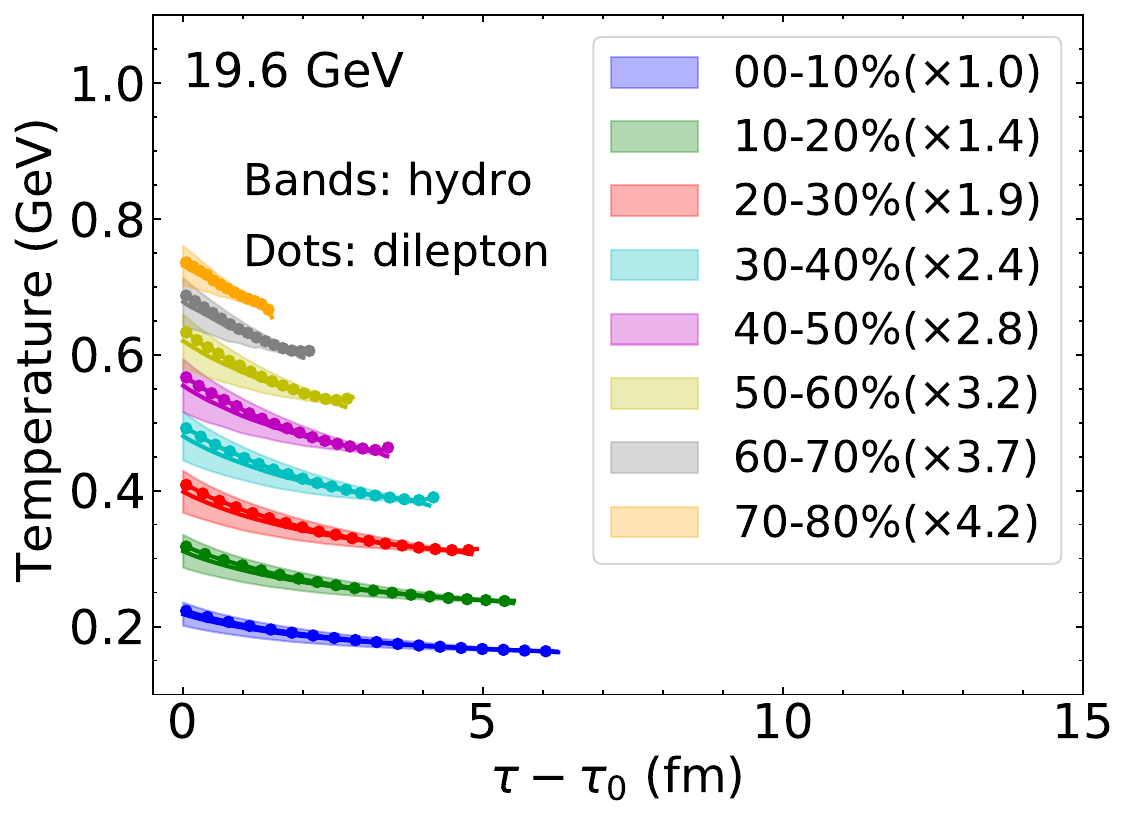}
\hspace{.5cm}
\includegraphics[width= .45\linewidth]{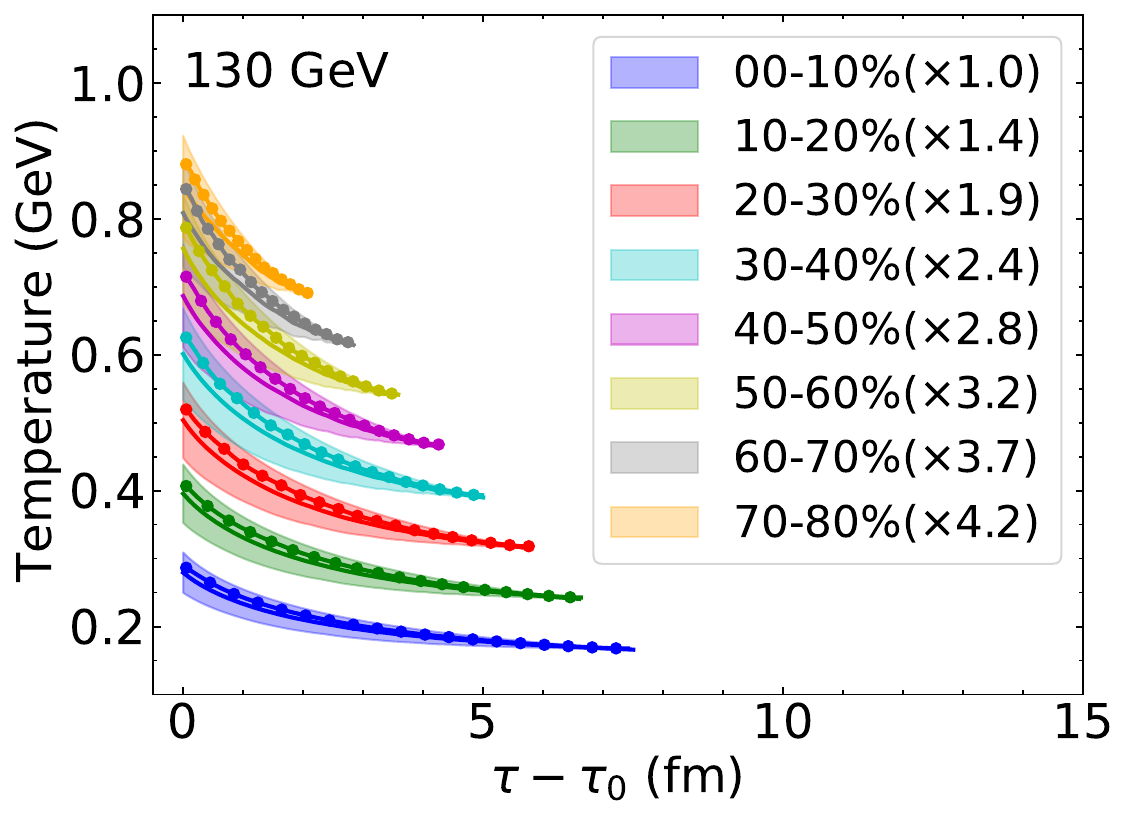}
\caption{%
        Comparison of effective temperatures derived from dilepton spectra (dots) and the hydrodynamic temperatures (bands) across various proper times, for Au+Au collisions within a range of centrality classes from 0--10\% to 70--80\% at two selected beam energies. To enhance clarity in the visualization, distinct multiplicative factors are applied to the temperatures corresponding to various centrality classes. The solid curve for hydrodynamic temperatures represents the mean values, with the band width indicating the standard deviations; see the texts for details.
    }
    \label{fig:temp_extraction_tau}
\end{center}
\end{figure*}

Figure~\ref{fig:temp_extraction_tau} depicts the comparison between effective temperature and hydrodynamic temperature as functions of the proper time for Au+Au collisions at two selected beam energies within centrality classes ranging from 0--10\% to 70--80\%. The dots interconnected by lines represent the effective temperatures \Teff{} derived from the dilepton spectra, while the curves accompanied by bands represent the mean value (\Tavg{}) and standard deviation ($\langle \delta T\rangle$) of hydrodynamic temperatures. The figure demonstrates a close alignment between \Teff{} and the hydrodynamic temperatures. 
The initial mean value (\Tin{}) and standard deviation ($\langle \delta T\rangle$) of temperatures, along with the corresponding proper time ($\tau_0$), when the hydrodynamic evolution starts for the eight beam energies across eight centrality classes, are listed in Table~\ref{tab:temperature-data}. In this work, the values of $\tau_0$ are obtained from Ref.~\cite{Shen:2020jwv}, initially motivated by the overlap time between the two colliding nuclei and further adjusted to reproduce the mean transverse momentum of identified particles \cite{Shen:2020jwv,Du:2023efk}.

\begin{table}[htbp]
\centering
\caption{Initial mean temperature (\Tin{}) and standard deviation ($\langle \delta T\rangle$), as well as effective temperature (\Teff{}), along with the initial proper time ($\tau_0$), for different beam energies and centrality classes.
}
\vspace{2mm}

\label{tab:temperature-data}
\begin{tabular}{cccllllllll}
\toprule\toprule
$\sqrt{\sNN}$ & $\tau_0$ & $T$ & \multicolumn{8}{c}{Centrality} \\
\cmidrule(lr){4-11}
\quad(GeV)\quad &\quad(fm)\quad & \quad(MeV)\quad\ & 0--10\% & 10--20\% & 20--30\% & 30--40\% & 40--50\% & 50--60\% & 60--70\% & 70--80\% \\
\midrule\midrule
\multirow{2}{*}{7.7} & \multirow{2}{*}{3.6} & $T_{\rm in}$ & 178 $\pm$ 12 & 175 $\pm$ 12 & 172 $\pm$ 12 & 169 $\pm$ 12 & 164 $\pm$ 11 & 159 $\pm$ 11 & 153 $\pm$ 11 & 144 $\pm$ 11 \\
& & \Teff{} & 173 $\pm$ 0 & 171 $\pm$ 0 & 170 $\pm$ 0 & 167 $\pm$ 0 & 165 $\pm$ 0 & 161 $\pm$ 0 & 157 $\pm$ 0 & 150 $\pm$ 0 \\
\midrule
\multirow{2}{*}{19.6} & \multirow{2}{*}{1.8} & $T_{\rm in}$ & 218 $\pm$ 17 & 214 $\pm$ 17 & 210 $\pm$ 16 & 204 $\pm$ 15 & 198 $\pm$ 14 & 191 $\pm$ 12 & 183 $\pm$ 10 & 176 $\pm$ 7 \\
& & \Teff{} & 199 $\pm$ 1 & 197 $\pm$ 1 & 194 $\pm$ 1 & 191 $\pm$ 1 & 187 $\pm$ 1 & 183 $\pm$ 0 & 177 $\pm$ 0 & 173 $\pm$ 0 \\
\midrule
\multirow{2}{*}{27} & \multirow{2}{*}{1.4} & $T_{\rm in}$ & 235 $\pm$ 21 & 231 $\pm$ 20 & 225 $\pm$ 20 & 219 $\pm$ 19 & 211 $\pm$ 18 & 202 $\pm$ 16 & 193 $\pm$ 13 & 184 $\pm$ 10 \\
& & \Teff{} & 209 $\pm$ 2 & 207 $\pm$ 2 & 204 $\pm$ 2 & 201 $\pm$ 2 & 196 $\pm$ 1 & 191 $\pm$ 1 & 184 $\pm$ 1 & 179 $\pm$ 0 \\
\midrule
\multirow{2}{*}{39} & \multirow{2}{*}{1.0} & $T_{\rm in}$ & 260 $\pm$ 26 & 254 $\pm$ 26 & 247 $\pm$ 25 & 239 $\pm$ 25 & 230 $\pm$ 23 & 219 $\pm$ 21 & 208 $\pm$ 18 & 196 $\pm$ 15 \\
& & \Teff{} & 223 $\pm$ 4 & 221 $\pm$ 3 & 218 $\pm$ 3 & 214 $\pm$ 3 & 209 $\pm$ 2 & 202 $\pm$ 2 & 195 $\pm$ 1 & 187 $\pm$ 1 \\
\midrule
\multirow{2}{*}{54.4} & \multirow{2}{*}{1.0} & $T_{\rm in}$ & 265 $\pm$ 27 & 259 $\pm$ 27 & 252 $\pm$ 26 & 244 $\pm$ 26 & 234 $\pm$ 24 & 223 $\pm$ 22 & 211 $\pm$ 19 & 198 $\pm$ 16 \\
& & \Teff{} & 227 $\pm$ 4 & 224 $\pm$ 4 & 221 $\pm$ 3 & 217 $\pm$ 3 & 211 $\pm$ 2 & 205 $\pm$ 2 & 197 $\pm$ 1 & 189 $\pm$ 1 \\
\midrule
\multirow{2}{*}{62.4} & \multirow{2}{*}{1.0} & $T_{\rm in}$ & 268 $\pm$ 27 & 262 $\pm$ 27 & 255 $\pm$ 27 & 246 $\pm$ 26 & 236 $\pm$ 25 & 225 $\pm$ 23 & 212 $\pm$ 19 & 200 $\pm$ 16 \\
& & \Teff{} & 229 $\pm$ 4 & 226 $\pm$ 4 & 223 $\pm$ 3 & 219 $\pm$ 3 & 213 $\pm$ 3 & 206 $\pm$ 2 & 198 $\pm$ 2 & 190 $\pm$ 1 \\
\midrule
\multirow{2}{*}{130} & \multirow{2}{*}{1.0} & $T_{\rm in}$ & 279 $\pm$ 30 & 273 $\pm$ 30 & 265 $\pm$ 29 & 256 $\pm$ 29 & 245 $\pm$ 27 & 233 $\pm$ 25 & 219 $\pm$ 21 & 204 $\pm$ 18 \\
& & \Teff{} & 236 $\pm$ 5 & 234 $\pm$ 4 & 230 $\pm$ 4 & 225 $\pm$ 4 & 219 $\pm$ 3 & 212 $\pm$ 2 & 203 $\pm$ 2 & 194 $\pm$ 1 \\
\midrule
\multirow{2}{*}{200} & \multirow{2}{*}{1.0} & $T_{\rm in}$ & 290 $\pm$ 32 & 283 $\pm$ 32 & 275 $\pm$ 31 & 265 $\pm$ 31 & 253 $\pm$ 29 & 240 $\pm$ 27 & 225 $\pm$ 24 & 210 $\pm$ 20 \\
& & \Teff{} & 244 $\pm$ 5 & 242 $\pm$ 5 & 238 $\pm$ 5 & 233 $\pm$ 4 & 226 $\pm$ 4 & 218 $\pm$ 3 & 209 $\pm$ 2 & 198 $\pm$ 1 \\
\midrule\midrule
\end{tabular}
\end{table}

\end{document}